\documentclass{article}
\usepackage{graphicx}
\usepackage{latexsym}
\usepackage{amsmath}
\usepackage{amssymb}
\usepackage{amsfonts}
\usepackage{amsxtra}
\usepackage[hypertex]{hyperref}

\allowdisplaybreaks
\newcommand{\sss}[1]{{\scriptscriptstyle #1}}

\newcommand{\Ll}{\ensuremath{\mathcal{L}}}

\newsavebox{\piracbox}
\savebox{\piracbox}{$\not\! p$}
\newcommand{\pirac}{\usebox{\piracbox}}
\newsavebox{\ptiracbox}
\savebox{\ptiracbox}{$\not\! p_\sss{T}$}
\newcommand{\ptirac}{\usebox{\ptiracbox}}
\newsavebox{\qiracbox}
\savebox{\qiracbox}{$\not\! q$}
\newcommand{\qirac}{\usebox{\qiracbox}}
\newcommand{\pt}{p_\sss{T}}
\newcommand{\dslash}[1]{\not\negmedspace #1\,}

\newcommand{\Tr}[1]{\mathrm{Tr}\left( #1 \right)}

\newcommand{\mm}{\widetilde{m}}
\newcommand{\mb}{m}
\newcommand{\Mb}{M_\chi}
\renewcommand{\O}{\mathcal{O}}
\newcommand{\T}{\mathcal{T}}
\newcommand{\G}{\mathcal{G}}
\newcommand{\F}{\mathcal{F}}
\newcommand{\A}{\mathcal{A}}
\renewcommand{\P}{\mathcal{P}}
\newcommand{\D}{\mathcal{D}}

\renewcommand{\Re}{\ensuremath{\mathrm{Re}}}
\renewcommand{\Im}{\ensuremath{\mathrm{Im}}}

\newcommand{\seq}{\doteq}
\newcommand{\umax}{u_\sss{\mathrm{max}}}
\newcommand{\umin}{u_\sss{\mathrm{min}}}
\newcommand{\ufmax}{u_{f_\mathrm{max}}}
\newcommand{\ufmin}{u_{f_\mathrm{min}}}
\newcommand{\uimax}{u_{i_\mathrm{max}}}
\newcommand{\uimin}{u_{i_\mathrm{min}}}
\newcommand{\tmax}{t_\sss{\mathrm{max}}}
\newcommand{\tmin}{t_\sss{\mathrm{min}}}
\newenvironment{sarray}{\renewcommand{\arraycolsep}{0pt}
  
  \begin{array}}{\end{array}} 
\newcommand{\uud}[4]{#1
  \begin{sarray}{ccc} {\scriptstyle #2} & {\scriptstyle #3} & \\ &  &
    {\scriptstyle #4} \end{sarray}} 
\newcommand{\udu}[4]{#1
  \begin{sarray}{ccc} {\scriptstyle #2} &  & {\scriptstyle #4}\\ &
    {\scriptstyle #3} &  \end{sarray}} 
\newcommand{\dud}[4]{#1
  \begin{sarray}{ccc} & {\scriptstyle #3}  & \\ {\scriptstyle #2} & &
    {\scriptstyle #4}  \end{sarray}} 
\newcommand{\udd}[4]{#1
  \begin{sarray}{ccc} {\scriptstyle #2} & & \\ & {\scriptstyle #3} &
    {\scriptstyle #4}  \end{sarray}} 
\newcommand{\ud}[3]{#1
  \begin{sarray}{cc}{\scriptstyle #2} &  \\ & {\scriptstyle #3}
  \end{sarray}} 
\title{Absorptive part of meson--baryon scattering amplitude and
  baryon polarization in chiral perturbation theory}
\author{Antonio O.\ Bouzas \thanks{E-mail:
    abouzas@mda.cinvestav.mx}\\\small Departamento de F\'{\i}sica
  Aplicada, CINVESTAV-IPN \\\small Carretera Antigua a Progreso Km.\
  6, Apdo.\ Postal 73 ``Cordemex''\\\small
  M\'erida 97310, Yucat\'an, M\'exico}
\date{}
\begin{document}
\maketitle
\begin{abstract}
  We compute the spin asymmetry and polarization of the final-state
  baryon in its rest frame in two-body meson--baryon low-energy
  scattering with unpolarized initial state, to lowest non-trivial
  order in BChPT.  The required absorptive amplitudes are obtained
  analytically at one-loop level.  We discuss the polarization results
  numerically for several meson--baryon processes.  Even at low
  energies above threshold, where BChPT can reasonably be expected to
  be applicable, sizable values of polarization are found for some
  processes. 
\end{abstract}
\section{Introduction}
\label{sec:intro}

Low-energy hadron dynamics is succesfully described by Chiral
Perturbation Theory (ChPT), the effective field theory of meson
strong, electromagnetic and weak interactions (see, \emph{e.g.,}
\cite{ber07,bor07,sch05} for recent reviews, \cite{wei96,don94} for
textbook expositions).  The effective chiral framework can also be
extended to the one-baryon sector, where a fully relativistic Baryon
Chiral Perturbation Theory (BChPT) has been formulated, describing
baryon interactions at low energies (recent reviews are given in
\cite{ber07a,sch05}. The equivalent heavy-baryon approach is reviewed,
\emph{e.g.,} in \cite{ber95}).  As in virtually all areas of hadronic
physics \cite{lea01}, in the low-energy regime spin phenomena are of
great interest as probes of the structure and dynamics of baryons.
Chiral effective theories are essential tools in the study of those
phenomena and have been applied, for instance, to compute nucleon spin
structure functions and spin-dependent polarizabilities \cite{ber07a}.

From the experimental point of view, spin observables at low- and
medium-energies have bee the subject of intensive studies in some
cases, but data is scarce in others.  There exists, for instance, a
substantial body of data on analyzing powers in nucleon--pion
scattering at energies around the $\Delta$ resonance peak (see
\cite{sevx,hofx1,hofx2} and refs.\ therein), and at lower energies
\cite{wiex,meix}.  In the strange sector data on spin observables at
low energies are definitely rare, though some exist~\cite{lovx}.

In this paper we consider a particular aspect of hadron spin dynamics,
the production of polarized baryons in unpolarized meson--baryon
scattering, in BChPT.  Such polarization effects are known to be large
in the high energy regime \cite{lea01,fel99} where, unlike at low
energy, the degrees of freedom relevant to the dynamics are the
partonic ones.  Here, we compute the spin asymmetry and polarization
of the final-state baryon in its rest frame in low-energy two-body
meson--baryon scattering with unpolarized initial state, to lowest
non-trivial order in BChPT.  Our main result is the expression for the
spin asymmetry to that order in closed analytical form.  From a more
qualitative point of view, our main motivation is to ascertain what
spin asymmetries and polarizations are obtained in BChPT, what
physical mechanisms are responsible for them, how they depend on the
kinematics of the scattering process (energy, scattering angle, and
flavor), and what is the order of magnitude of the polarization
effects obtained.

The paper is organized as follows.  In the following section we
present our notation and conventions, including our choice of flavor
basis, the parameterization for the scattering amplitude and the
related expressions for the spin asymmetry and polarization, and the
tree-level amplitudes.  The absorptive part of the amplitude is
obtained analytically at one-loop level in Sect.\ \ref{sec:absptv}.
This result completes the calculation of the asymmetry and
polarization at lowest non-trivial order in BChPT.  In Sect.\
\ref{sec:discu} we discuss our results in a more concrete setting,
illustrating them with numerical computations of the polarization for
several meson--baryon processes.  In Sect.\ \ref{sec:final} we give
some final remarks.

\section{Scattering amplitude and polarization} 
\label{sec:preli}

The ground-state meson and baryon octets are described by standard
\cite{don94} traceless $3\times 3$ complex matrix fields $\phi$ and
$B$, resp., with $\phi$ hermitian.  We use the physical flavor basis
\begin{equation}
  \label{eq:cartanweyl}
  \begin{gathered}
    \beta^1 = \frac{1}{\sqrt{2}} \left(\lambda^1 + i \lambda^2\right),
    \quad
    \beta^1 = \beta^{2\dagger}~,
    \quad
  \beta^3 = \lambda^3~,\\
  \beta^4 = \frac{1}{\sqrt{2}} \left(\lambda^4 + i \lambda^5\right),
  \quad
  \beta^5 = \beta^{4\dagger}~,\quad  
  \beta^6 = \frac{1}{\sqrt{2}} \left(\lambda^6 + i \lambda^7\right),\quad
  \beta^7 = \beta^{6\dagger}~,\quad  
  \beta^8 = \lambda^8~,
  \end{gathered}
\end{equation}
where $\lambda^a$ are SU(3) Gell-Mann matrices. The real matrices
$\beta^a$ are not hermitian.  Their hermitian conjugates form a basis
that differs from $\{\beta^a\}_{a=1}^8$ only in its ordering.  To
distinguish field components with respect to each of those bases we
use lower flavor indices for ${\beta^\dagger}_a$.  Thus, meson and
baryon fields are decomposed as\footnote{We do not use summation
  convention for flavor indices.}  $\phi = \sum_b \phi_b \beta^b
/\sqrt{2} = \sum_b \phi^b {\beta^\dagger}_b /\sqrt{2}$ and $B = \sum_a
B_a \beta^a /\sqrt{2} = \sum_a B^a {\beta^\dagger}_a/\sqrt{2}$, with
$\phi_b = \Tr{{\beta^\dagger}_b \phi}/\sqrt{2}$, $\phi^b = \Tr{\beta^b
  \phi}/\sqrt{2}$, and similarly $B_a$ and $B^a$.  Baryon and meson
states are denoted $|B^a(p,\sigma)\rangle$ and $|M^b(q)\rangle$,
resp., with $\sigma=\pm 1/2$ the spin along a fixed spatial direction
in the fermion rest frame, and $p$, $q$ four-momenta.  We always use
hadron kets with an upper index, and bras with a lower one, with
masses $\mb_a$ and $\mm_b$ for baryons and mesons resp.  Free fields
couple to one-particle states as $\langle 0 | B_a(x) |
B^c(p,\sigma)\rangle = \delta_a^c u(p,\sigma) \exp(-ipx)$ and $\langle
0 | \phi_b(x) | M^c(q)\rangle = \delta_b^c \exp(-iqx)$.

Indices can be raised or lowered by means of the symmetric matrices
$e^{ab}=\Tr{\beta^a\beta^b}/2$ $=\Tr{{\beta^\dagger}_a
  {\beta^\dagger}_b}/2$ $= e_{ab}$ and $e^a_b = \delta^a_b$.  In this
basis the structure constants and the anticommutator constants in the
fundamental representation are
\begin{equation}
  \label{eq:fand}
  \begin{aligned}\mbox{ }
    [\beta^a,\beta^b] &= 2 \sum_c \uud{f}{a}{b}{c}\, \beta^c~,
    & & \uud{f}{a}{b}{c} = \frac{1}{4} \Tr{{\beta^\dagger}_c[\beta^a,
      \beta^b]},\\ 
    \{\beta^a,\beta^b\} &= 2 \sum_c \uud{d}{a}{b}{c}\, \beta^c +
    \frac{4}{3} e^{ab}~,
    & & \uud{d}{a}{b}{c} = \frac{1}{4} \Tr{{\beta^\dagger}_c\{\beta^a,
      \beta^b\}}.
  \end{aligned}  
\end{equation}
Similar definitions hold for $\dud{f}{a}{b}{c\,}$,
$\udd{d}{a}{b}{c}\,$, etc.  The constants $f^{abc}$ and $d^{abc}$ (as
well as $f_{abc} = - f^{abc}$ and $d_{abc} = d^{abc}$) are totally
antisymmetric and symmetric, resp.  Their numerical values are
different from their Gell-Mann--basis analogs. 

With the $T$-matrix defined in terms of the $S$-matrix as $S=I + i
(2\pi)^4 \delta(P_f-P_i) T$, the scattering amplitudes are given by
$T$-matrix elements $\ud{\T}{ab}{a'b'}(s,u;\sigma,\sigma') \equiv
\langle B_{a'}(p',\sigma') M_{b'}(q') | T | B^{a}(p,\sigma) M^b(q)
\rangle$ as functions of the Mandelstam invariants $s=(p+q)^2$, $u =
(p-q')^2$ and the spin variables.  The most general form for the
two-body meson-baryon scattering amplitude consistent with Lorentz
invariance and the discrete symmetries of the strong interactions is
\begin{subequations}
\label{eq:paramet}
\begin{equation}
\label{eq:parameta}
\ud{\T}{ab}{a'b'} =
\overline{u}'(p',\sigma')\ud{\Gamma}{ab}{a'b'}\,u(p,\sigma), 
\qquad
\ud{\Gamma}{ab}{a'b'} = \ud{\Gamma_0}{ab}{a'b'} +
  \ud{\Gamma_1}{ab}{a'b'} \ptirac~,
\end{equation}
with $\pt$ the total momentum, and $\ud{\Gamma}{ab}{a'b'}$ depending
only on $s$, $u$.  Below we will make repeated use of the unitarity
relation for two-body scattering amplitudes, involving an integration
volume element depending on $\pt$, which justifies our choice of a
parameterization of $\ud{\Gamma}{ab}{a'b'}$ in terms of $\ptirac$.
Complex conjugation of the first equality in (\ref{eq:parameta})
yields the matrix element $\ud{\T^\dagger}{a'b'}{ab} = \langle
B_{a}(p,\sigma) M_{b}(q) | T^\dagger | B^{a'}(p',\sigma') M^{b'}(q')
\rangle$ written as,
\begin{equation}
\label{eq:parametb}
  \ud{\T^\dagger}{a'b'}{ab} =
  \overline{u}(p,\sigma)\ud{\overline\Gamma}{a'b'}{ab}\,u'(p',\sigma'),   
  \qquad
  \ud{\overline\Gamma}{a'b'}{ab} = \ud{\overline\Gamma_0}{a'b'}{ab} +
  \ud{\overline\Gamma_1}{a'b'}{ab} \ptirac = 
  (\ud{\Gamma_0}{ab}{a'b'})^* +
  (\ud{\Gamma_1}{ab}{a'b'})^* \ptirac~.
\end{equation}
\end{subequations}
The dispersive and absorptive parts of the amplitude are then
$\ud{\Gamma}{ab}{a'b'} = \ud{\Gamma_D}{ab}{a'b'} + i
\ud{\Gamma_A}{ab}{a'b'}$ with,
\begin{equation}
  \label{eq:disabs}
  \begin{aligned}
    \ud{\Gamma_D}{ab}{a'b'} &= \frac{1}{2} \left(
      \ud{\Gamma}{ab}{a'b'} + \ud{\overline\Gamma}{a'b'}{ab}\right)
    \seq \Re\left( \ud{\Gamma_0}{ab}{a'b'} \right) +
    \Re\left( \ud{\Gamma_1}{ab}{a'b'} \right) \ptirac~,
    \\
    \ud{\Gamma_A}{ab}{a'b'} &= \frac{1}{2i} \left(
      \ud{\Gamma}{ab}{a'b'} - \ud{\overline\Gamma}{a'b'}{ab}\right)
    \seq \Im\left( \ud{\Gamma_0}{ab}{a'b'} \right) +
    \Im\left( \ud{\Gamma_1}{ab}{a'b'} \right) \ptirac~.
  \end{aligned}
\end{equation}
In this equation, and in what follows, the symbol ``$\seq$'' means
that equality holds when both sides are sandwiched between
$\overline{u}'(p',\sigma')$ and $u(p,\sigma)$.  Although time-reversal
invariance is already taken into account in the Dirac structure of
(\ref{eq:paramet}), it also implies restrictions on the flavor
dependence of the amplitude.  In the flavor basis
(\ref{eq:cartanweyl}), time-reversal invariance is expressed as
\begin{equation}
  \label{eq:timerev}
  \overline{u}'(p',\sigma')\left( \ud{\Gamma_0}{ab}{a'b'} +
  \ud{\Gamma_1}{ab}{a'b'} \ptirac \right)u(p,\sigma) =  
  \overline{u}'(p',\sigma') \left( \ud{\Gamma_0}{a'b'}{ab} +
  \ud{\Gamma_1}{a'b'}{ab} \ptirac \right) u(p,\sigma),
\end{equation}
where we again omitted the arguments $s$, $u$, which are invariant
under time inversion.  

The Dirac spinor $u'(p',\sigma')$ satisfies the polarization equation
$\gamma_5 \dslash{s}' u'(p',\sigma') = -2\sigma' u'(p',\sigma')$,
where $s'^\mu$ is the spin four-vector defined by the condition that
in the fermion rest frame $s'^\mu = (0,\hat{s}'_*)$ with $\hat{s}'_*$
the versor lying on the fixed spin-quantization axis.  The following
relations then hold, $s'^\mu s'_\mu = -1$, $s'^\mu p'_\mu = 0$, and
\begin{equation}
  \label{eq:spin}
  u'(p',\sigma') \overline{u}'(p',\sigma') = \frac{1}{2} (\pirac' +
  \mb_{a'}) (1- 2\sigma' \gamma_5 \dslash{s}').
\end{equation}
The total unpolarized squared amplitude, symbolically denoted
$|\T|^2$, is given by
\begin{subequations}
  \label{eq:totsqam}
  \begin{gather}
  \begin{aligned}
    |\T|^2  & \equiv \frac{1}{2} \sum_{\sigma,\sigma'} \ud{\T}{ab}{a'b'}
    \ud{\T^\dagger}{a'b'}{ab}
    =
  \frac{1}{2} \Tr{\ud{\Gamma}{ab}{a'b'} (\pirac + \mb_a)
    \ud{\overline\Gamma}{a'b'}{ab} (\pirac' + \mb_{a'})}
  \\
  &= C_{00} \left|\ud{\Gamma_0}{ab}{a'b'} \right|^2 +
  C_{11} \left|\ud{\Gamma_1}{ab}{a'b'} \right|^2 + 2 C_{01}
  \Re\left(\ud{\Gamma_0}{ab}{a'b'} \ud{\Gamma_1}{ab}{a'b'}^* \right),
  \end{aligned}  \label{eq:totsqam1}\\
\intertext{with}
  \begin{gathered}
    C_{00} =  ((\mb_a + \mb_{a'})^2 - t),
    \qquad
    C_{01} = \mb_a (s+\mb_{a'}^2 - \mm_{b'}^2) +
    \mb_{a'} (s+\mb_{a}^2-\mm_b^2),
    \\
    C_{11} = (s+\mb_{a'}^2 - \mm_{b'}^2) (s+\mb_{a}^2 -
    \mm_{b}^2) + s \left(t-(\mb_{a'}-\mb_a)^2\right).
  \end{gathered}\label{eq:totsqam2}
  \end{gather}
\end{subequations}
From (\ref{eq:paramet}) and (\ref{eq:spin}) the spin asymmetry $\A$
for unpolarized two-body scattering is written as,
\begin{equation}
  \label{eq:asym}
  \begin{aligned}
  \A  & \equiv \frac{1}{2} \sum_{\sigma,\sigma'}
  (2 \sigma') \ud{\T}{ab}{a'b'} \ud{\T^\dagger}{a'b'}{ab} = 
  \frac{1}{2} \Tr{\ud{\Gamma}{ab}{a'b'} (\pirac + \mb_a)
    \ud{\overline\Gamma}{a'b'}{ab} (\pirac' + \mb_{a'}) \dslash{s}'
    \gamma_5} 
  \\
  & = -i \left( \Re\left( \ud{\Gamma_0}{ab}{a'b'} \right)
    \Im\left( \ud{\Gamma_1}{ab}{a'b'} \right) -
    \Im\left( \ud{\Gamma_0}{ab}{a'b'} \right)
    \Re\left( \ud{\Gamma_1}{ab}{a'b'} \right)\right)
  \Tr{\pirac \ptirac \pirac' \dslash{s}' \gamma_5}.
  \end{aligned}
\end{equation}
Since
\begin{equation}
  \label{eq:spina}
\begin{aligned}
  -i\Tr{\pirac \ptirac \pirac' \dslash{s}' \gamma_5}
  &=
  4 \varepsilon^{\alpha\beta\gamma\delta} p_\alpha {\pt}_\beta 
  p'_\gamma s'_\delta \\
&= 4 \mb_{a'} \sqrt{\vec{p}\,^2 \vec{q}\,^2 - (\vec{p}\cdot
  \vec{q})^2} \;  \hat{s}'_*\cdot \hat{n}~,
\qquad
\hat{n} = \frac{\vec{q}\wedge\vec{p}}{|\vec{q}\wedge\vec{p}\,|}~,
\qquad\qquad
\text{(lab$'$)}
\end{aligned}
\end{equation}
where the second equality holds in the rest frame of the final baryon
(henceforth ``lab$'$'' frame), we see that for an unpolarized initial
state the final-state spin asymmetry in the lab$'$ frame lies along
the direction $\hat{n}$ orthogonal to the plane of the reaction.  We
denote $\A'_*$ the asymmetry $\A$ expressed in lab$'$ frame, as in
(\ref{eq:spina}), and with $\hat{s}'_* = \hat{n}$.  The polarization
of the final baryon along the direction $\hat{n}$ in its rest frame is
then
\begin{equation}
  \label{eq:polariz}
  \P'_* = \frac{\A'_*}{|\T|^2}~.
\end{equation}
We see from the second line of (\ref{eq:asym}) that $\A'_*$ and
$\P'_*$ vanish unless real particles are created in intermediate
states.  Indeed, if only virtual--particle intermediate states are
possible then the imaginary parts $\Im(\Gamma_{0,1}) = 0$ in
(\ref{eq:asym}), leading to a vanishing asymmetry. This occurs, in
particular, at tree level. To obtain $\A'_*$ and $\P'_*$ to lowest
non-trivial order we must compute the dispersive part of the amplitude
at tree level, and its absorptive part at one loop.

\subsection{Tree-level amplitudes}
\label{sec:lag}

The Lagrangian of fully relativistic Baryon Chiral Perturbation Theory
(BChPT) is written as a sum $\Ll = \Ll_M + \Ll_{MB}$ of a purely
mesonic Lagrangian $\Ll_M$ and a meson--baryon one $\Ll_{MB}$.  The
mesonic Lagrangian to $\mathcal{O}(q^4)$\footnote{ $\mathcal{O}(q^n)$
  denotes a generic quantity of chiral order $n$, with $q$ a nominally
  small quantity such as a meson momentum or mass.} was first obtained
in \cite{gas84,gas85a}.  The meson--baryon Lagrangian $\Ll_{MB}$ has
been given to $\mathcal{O}(q^3)$ in the three-flavor case in
\cite{kra90,fri04,oll06a} (and in \cite{gas88} for two flavors).  Here
we discuss the tree-level amplitudes for meson--baryon scattering
obtained from $\Ll$.

In the parameterization (\ref{eq:parameta}) it is convenient to write
$\ud{\T}{ab}{a'b'}$ as a sum over Feynman graphs $\G$, explicitly
factoring the flavor coefficients from interaction vertices,
\begin{equation}
  \label{eq:graphs}
  \ud{\Gamma}{ab}{a'b'} \seq \ud{\Gamma_0}{ab}{a'b'} +
  \ud{\Gamma_1}{ab}{a'b'} \ptirac  = \sum_\G \sum_{\{I\}_\G}
  \udu{\F_{(\G)}}{ab}{a'b'}{\{I\}} \left( \widehat{\Gamma}_{(\G)0}^{\{I\}} +
    \widehat{\Gamma}_{(\G)1}^{\{I\}} \ptirac \right) .
\end{equation}
The index $\G$ in (\ref{eq:graphs}) runs over all Feynman graphs
contributing to the amplitude $\ud{\T}{ab}{a'b'}$.  The second sum
runs over a set $\{I\}_\G$ of internal flavor indices for each diagram
$\G$.  In the flavor limit the reduced form factors
$\widehat{\Gamma}_{(\G)0,1}^{\{I\}}$ carry only internal flavor
indices but, in fact, they also depend on $a$, $a'$ through their
dependence on initial and final baryon masses. At tree level $\G$
takes the values $c$ (contact-interaction diagram), $s$ and $u$.  The
set $\{I\}_{c}$ is empty, and each of $\{I\}_{s,u}$ contains only one
flavor index originating in the mass of the internal baryon
line.\footnote{Those flavor indices are arbitrarily written as upper
  indices, but between braces to denote their non-tensorial nature.}
The tree-level amplitude $\ud{\T}{ab}{a'b'}$ can then be written as,
\begin{subequations}
  \label{eq:tree}
  \begin{gather}
    \ud{\F_{(c)}}{ab}{a'b'} = \sum_d \dud{f}{b'}{b}{d}
    \uud{f}{d}{a}{a'} ~,\quad
    \widehat{\Gamma}_{(c)0} = -\frac{1}{2f^2}(\mb_a+\mb_{a'})~,\quad
    \widehat{\Gamma}_{(c)1} = \frac{1}{f^2}~,
    \\
    \begin{gathered}
    \udu{\F_{(s)}}{ab}{a'b'}{\{d\}} =
    \left( D\uud{d}{b}{a}{d}   + F\uud{f}{b}{a}{d}
      \rule{0pt}{10pt}\right)
    \left( D\udd{d}{d}{b'}{a'} - F\udd{f}{d}{b'}{a'}\right) ,\quad
    \widehat{\Gamma}_{(s)1}^{\{d\}} =
    -\frac{1}{f^2}\left(1 +
      \frac{(\mb_{a'}+\mb_d)(\mb_a+\mb_d)}{s-\mb_d^2}\right),     
    \\
    \widehat{\Gamma}_{(s)0}^{\{d\}} =
    \frac{1}{f^2}\left(\mb_a+\mb_{a'}+\mb_d+
    \mb_d \frac{(\mb_{a'}+\mb_d)(\mb_a+\mb_d)}{s-\mb_d^2}\right), 
    \end{gathered}
    \\
    \begin{gathered}
    \udu{\F_{(u)}}{ab}{a'b'}{\{d\}} =
    \left( D\udd{d}{a}{b'}{d}   - F\udd{f}{a}{b'}{d}\right)
    \left( D\uud{d}{b}{d}{a'} + F\uud{f}{b}{d}{a'}\right) ,\quad
    \widehat{\Gamma}_{(u)1}^{\{d\}} =
    \frac{1}{f^2}\left(1 +
      \frac{(\mb_{a'}+\mb_d)(\mb_a+\mb_d)}{u-\mb_d^2}\right),     
    \\
    \widehat{\Gamma}_{(u)0}^{\{d\}} =
    \frac{1}{f^2}\left(\mb_d - (\mb_{a'}+\mb_a-\mb_d)
    \frac{(\mb_{a'}+\mb_d)(\mb_a+\mb_d)}{u-\mb_d^2}\right). 
    \end{gathered}
  \end{gather}
\end{subequations}
Here $f$ is the pseudoscalar-meson decay constant in the chiral limit.
The coupling constants $D$ and $F$ have been obtained from
experimental data on hyperon semileptonic decays in
\cite{clo93,bor99,rat99}.  The amplitudes (\ref{eq:tree}) are obtained
by resumming mass terms from the $\O(q^2)$ Lagrangian and
incorporating them into the $\O(q^1)$ free baryon Lagrangian,
thereby explicitly taking into account baryon mass splittings.  This
procedure is not inconsistent at leading order in the chiral expansion
provided the result is not used as the basis of a next-to-leading
order calculation.  The purely leading-order result, however, can be
trivially recovered from (\ref{eq:tree}) by setting $\mb_a=\Mb$
for all $a=1,\ldots,8$ in our expressions. 

If we set in (\ref{eq:tree}) all baryon masses to their chiral-limit
common value $\Mb$, expand the Dirac spinors
$\overline{u}'(p',\sigma')$ and $u(p,\sigma)$ into their upper and
lower components, and change the flavor basis to the Gell-Mann basis,
we recover the tree-level amplitudes of \cite{oll01}.  Furthermore,
for those flavor coefficients considered in \cite{ose98,jid02} we
find full numerical agreement with their tabulated values.

\section{One-loop absorptive parts}
\label{sec:absptv}

The spin asymmetry (\ref{eq:asym}) involves the absorptive part of the
amplitude, which vanishes at tree level for physical values of the
external momenta.  At one-loop level the only diagrams that contribute
to the absorptive part in the physical region are those that can be
factored as products of tree-level diagrams with on-shell external
legs \cite{col65}.  From (\ref{eq:paramet}), (\ref{eq:disabs}),
(\ref{eq:timerev}) we have $\overline{u}'(p',\sigma')
\ud{\Gamma_A}{ab}{a'b'}\, u(p,\sigma) = 1/(2i) \langle
B_{a'}(p',\sigma') M_{b'}(q') | (T - T^\dagger) | B^{a}(p,\sigma)
M^b(q) \rangle $.  Hence, from the unitarity relation $T-T^\dagger = i
(2\pi)^4 \delta(P_f - P_i) TT^\dagger$ we obtain the cutting rules for
one-loop diagrams,
\begin{multline}
  \label{eq:cutting}
  \overline{u}'(p',\sigma') \ud{\Gamma_A}{ab}{a'b'}\, u(p,\sigma) =
  \frac{1}{8\pi^2} \int d^4Q d^4R\,
  \delta(\pt -Q -R) \sum_{h,h'} \delta_+(Q^2 - \mm_{h'}^2)
  \delta_+(R^2 - \mb_{h}^2) 
  \\
  \times \overline{u}'(p',\sigma') \ud{\Gamma}{hh'}{a'b'}
  (\dslash{R}+\mb_h) \ud{\overline{\Gamma}}{ab}{hh'}\, u(p,\sigma).
\end{multline}
A diagrammatic representation of this equation is given in fig.\
\ref{fig:fig1}.  Notice that meson--meson vertices do not enter the
absorptive part at this order.  We introduce the notation 
\begin{equation}
\label{eq:cutting2}   
 \int dV_R\, \left( \widehat{\Gamma}_{(\G')0}^{\{d'\}} + 
    \widehat{\Gamma}_{(\G')1}^{\{d'\}} \ptirac \right)
  (\dslash{R} + \mb_h)   \left( \widehat{\Gamma}_{(\G)0}^{\{d\}} +
    \widehat{\Gamma}_{(\G)1}^{\{d\}} \ptirac \right) 
  \seq   \left( \widehat{\Gamma}_{(\G\G')0}^{\{dd'\}} +
  \widehat{\Gamma}_{(\G\G')1}^{\{dd'\}} \ptirac \right),
\end{equation}
where $\G, \G' = c, s, u$ and the integration measure $dV_R$, whose
dependence on $\mb_{h}$, $\mm_{h'}$ is not explicitly shown in the
notation, is defined in (\ref{eq:volume}).  Similarly, the dependence
of $\widehat{\Gamma}_{(\G')0,1}^{\{d'\}}$ on $\mb_{a'}$, $\mb_{h}$ and
that of $\widehat{\Gamma}_{(\G)0,1}^{\{d\}}$ on $\mb_{a}$, $\mb_{h}$
are not explicitly indicated. More specifically, the reduced
amplitudes $\widehat{\Gamma}_{(\G')0,1}^{\{d'\}}$ and
$\widehat{\Gamma}_{(\G)0,1}^{\{d\}}$ in (\ref{eq:cutting2}) correspond
to the amplitudes $\langle B_{a'}(p',\sigma')
M_{b'}(q')|T|B^{h}(R,\Sigma) M^{h'}(Q)\rangle$ and $\langle
B_{h}(R,\Sigma) M_{h'}(Q)|T|B^{a}(p,\sigma) M^{b}(q)\rangle$, resp.,
evaluated at tree level.  They are given by (\ref{eq:tree}), with the
appropriate changes in flavor, momentum, and spin variables.  With
(\ref{eq:graphs}) and (\ref{eq:cutting2}), we can rewrite
(\ref{eq:cutting}) in the more compact form
\begin{equation}
\label{eq:cutting3}
  \ud{\Gamma_A}{ab}{a'b'} \seq \sum_{\G,\G'}
  \sum_{h,h',d,d'}  \udu{\F_{(\G')}}{hh'}{a'b'}{\{d'\}}
  \udu{\F_{(\G)}}{ab}{hh'}{\{d\}} \frac{1}{8\pi^2}
  \left( \widehat{\Gamma}_{(\G\G')0}^{\{dd'\}} +
  \widehat{\Gamma}_{(\G\G')1}^{\{dd'\}} \ptirac
\right),  
\end{equation}
In (\ref{eq:cutting2}) and (\ref{eq:cutting3}), if $\G$ or $\G'=c$ the
respective superindex $d$ or $d'$ must be omitted, since there are no
internal particles propagating in the contact diagram.

\subsection{Bubble diagrams} 
\label{sec:bubble}

In those terms on the r.h.s.\ of (\ref{eq:cutting3}) not involving
$\G$ or $\G'=u$ the reduced form factors $\widehat{\Gamma}_{(\G)0,1}$
and $\widehat{\Gamma}_{(\G')0,1}$ are independent of $R^\mu$, so the
integration is trivial.  For those terms we have
\begin{equation}
  \label{eq:cands}
  \begin{aligned}
    \widehat{\Gamma}_{(\G\G')0}^{\{dd'\}} &= s B_1^{\{hh'\}}
    H_{(\G\G')}^{\{dd'\}} + \mb_h B_0^{\{hh'\}} K_{(\G\G')}^{\{dd'\}}~,
    &  
    \widehat{\Gamma}_{(\G\G')1}^{\{dd'\}} &= \mb_h B_0^{\{hh'\}}
    H_{(\G\G')}^{\{dd'\}} + B_1^{\{hh'\}} K_{(\G\G')}^{\{dd'\}}~,
    \\
    H_{(\G\G')}^{\{dd'\}} &= \widehat{\Gamma}_{(\G')0}^{\{d'\}}
    \widehat{\Gamma}_{(\G)1}^{\{d\}} + \widehat{\Gamma}_{(\G')1}^{\{d'\}}
    \widehat{\Gamma}_{(\G)0}^{\{d\}}~,
    &  
    K_{(\G\G')}^{\{dd'\}} &= \widehat{\Gamma}_{(\G')0}^{\{d'\}}
    \widehat{\Gamma}_{(\G)0}^{\{d\}} + s \widehat{\Gamma}_{(\G')1}^{\{d'\}}
    \widehat{\Gamma}_{(\G)1}^{\{d\}}~,  
  \end{aligned}
\end{equation}
with $\G,\G'=c,s$. The phase-space integrals $B_{0,1}^{\{hh'\}}$ are
given in \ref{sec:integrals}. The quantities
$H_{(\G\G')}^{\{dd'\}}$ and $K_{(\G\G')}^{\{dd'\}}$ are introduced
here for notational convenience.  As with
$\widehat{\Gamma}_{(\G)0,1}^{\{d\}}$ and
$\widehat{\Gamma}_{(\G')0,1}^{\{d'\}}$, their dependence on $a$, $a'$,
$h$, $h'$ through initial and final state masses is not indicated
explicitly for simplicity.

From (\ref{eq:tree}) and the second line of (\ref{eq:cands}) we have, 
\begin{equation}  \label{eq:candshank}
\begin{array}{rclcrcl}
f^4 H_{(cc)} & = & \displaystyle -\frac{1}{2} (\mb_a+\mb_{a'}+2\mb_h)~,  
&\hspace{60pt}&
f^4 K_{(cc)} & = & \displaystyle s+\frac{1}{4} (\mb_a
+ \mb_h) (\mb_{a'} + \mb_h), 
\\[10pt]
f^4 H_{(cs)}^{\{d'\}} & = & \multicolumn{5}{l}{\displaystyle
(\mb_{a' }+ \mb_h) +
\frac{1}{2} (\mb_a + \mb_h + 2\mb_{d'})
\left( 1+\frac{(\mb_{a'} + \mb_{d'}) (\mb_{h} +
\mb_{d'})}{s-\mb_{d'}^2}
\right),} 
\\[10pt]
f^4 K_{(cs)}^{\{d'\}} & = & \multicolumn{5}{l}{\displaystyle
-\frac{1}{2}(\mb_{a' }+ \mb_h) (\mb_{a}+ \mb_h) - \left(s +
\frac{\mb_{d'}}{2}(\mb_a + \mb_h)\right) \left( 1+\frac{(\mb_{a'} +
    \mb_{d'}) 
(\mb_{h} + \mb_{d'})}{s-\mb_{d'}^2}\right),}
\\[10pt]
H_{(sc)}^{\{d\}} & = &\left. H_{(cs)}^{\{d\}}
\right|_{a\leftrightarrow a'}~,
& &
K_{(sc)}^{\{d\}} & = &
\left. K_{(cs)}^{\{d\}} \right|_{a\leftrightarrow a'}~, 
\\[10pt]
f^4 H_{(ss)}^{\{dd'\}} & = & \multicolumn{5}{l}{\displaystyle 
  -\left( 1+\frac{(\mb_{h} + \mb_{d}) (\mb_{a} +
\mb_{d})}{s-\mb_{d}^2}\right) (\mb_{a'} + \mb_h) - 
\left( 1+\frac{(\mb_{h} + \mb_{d'})
(\mb_{a'} + \mb_{d'})}{s-\mb_{d'}^2}\right)}
\\[10pt]
 & \multicolumn{6}{l}{\displaystyle
\times (\mb_{a} + \mb_h)  - 
(\mb_d + \mb_{d'}) \left( 1+\frac{(\mb_{h} + \mb_{d})
(\mb_{a} + \mb_{d})}{s-\mb_{d}^2}\right)
\left( 1+\frac{(\mb_{h} + \mb_{d'}) (\mb_{a'} +
    \mb_{d'})}{s-\mb_{d'}^2}\right)},
\\[10pt]
f^4 K_{(ss)}^{\{dd'\}} & = & \multicolumn{5}{l}{\displaystyle
  (\mb_h + \mb_{a'}) (\mb_h + \mb_{a})
+\mb_d (\mb_h + \mb_{a'}) \left( 1 + \frac{(\mb_{h} + \mb_{d}) 
(\mb_{a} +  \mb_{d})}{s-\mb_{d}^2}\right)} 
\\[10pt]
& \multicolumn{6}{l}{\displaystyle
 + \mb_{d'} (\mb_h + \mb_{a})
          \left( 1+\frac{(\mb_{h} + \mb_{d'}) (\mb_{a'} +
              \mb_{d'})}{s-\mb_{d'}^2}\right)}
\\[10pt]
& \multicolumn{6}{l}{\displaystyle
 + (s + \mb_{d} \mb_{d'})
      \left( 1+\frac{(\mb_{h} + \mb_{d}) (\mb_{a} +
          \mb_{d})}{s-\mb_{d}^2}\right) \left( 1+\frac{(\mb_{h} +
          \mb_{d'}) (\mb_{a'} + \mb_{d'})}{s-\mb_{d'}^2}\right).}
  \end{array}
\end{equation}
The contributions to the absorptive part (\ref{eq:cutting3}) of the
one-loop amplitude from diagrams $(cc)$, $(cs)$, $(sc)$, $(ss)$ (see
fig.\ \ref{fig:fig1}) is then given by (\ref{eq:cands}) and
(\ref{eq:candshank}).

\subsection{Triangle diagrams}
\label{sec:triangle}

The terms in (\ref{eq:cutting3}) for which $\G'=u\neq\G$ and those
with $\G'\neq u=\G$ are almost identical.  We will consider the former
case first and then apply the results to the latter.  For the diagrams
$(cu)$, $(su)$ in fig.\ \ref{fig:fig1} the last factor in the
integrand in (\ref{eq:cutting2}) is independent of $R^\mu$.  Their
contribution to (\ref{eq:cutting3}) is then of the form
\begin{equation}
  \label{eq:cu3}
  \widehat{\Gamma}_{(\G u)0}^{\{dd'\}} = \Omega_{(u_f)0}^{\{d'\}}
  \widehat{\Gamma}_{(\G)0}^{\{d\}} + s \Omega_{(u_f)1}^{\{d'\}}
  \widehat{\Gamma}_{(\G)1}^{\{d\}}~,
  \qquad
  \widehat{\Gamma}_{(\G u)1}^{\{dd'\}} = \Omega_{(u_f)0}^{\{d'\}}
  \widehat{\Gamma}_{(\G)1}^{\{d\}} + \Omega_{(u_f)1}^{\{d'\}}
  \widehat{\Gamma}_{(\G)0}^{\{d\}}~, 
\end{equation}
with $\Omega_{(u_f)0,1}^{\{d'\}}$ defined by the relation 
\begin{equation}
  \label{eq:cu4}
    \int dV_R \left(\widehat{\Gamma}_{(u)0}^{\{d'\}} +
    \widehat{\Gamma}_{(u)1}^{\{d'\}} \ptirac \right)
  (\dslash{R}+\mb_h) \seq
   \Omega_{(u_f)0}^{\{d'\}} +
  \Omega_{(u_f)1}^{\{d'\}} \ptirac~.
\end{equation}
In order to obtain $\Omega_{(u_f)0,1}^{\{d'\}}$ we further split
(\ref{eq:cu4}) as
\begin{subequations}
  \label{eq:cu5}
  \begin{align}
  \Omega_{(u_f)0}^{\{d'\}} &= \Omega_{(u_f)00}^{\{d'\}} +
  \Omega_{(u_f)10}^{\{d'\}}~,
  &
  \Omega_{(u_f)1}^{\{d'\}} &= \Omega_{(u_f)01}^{\{d'\}} +
  \Omega_{(u_f)11}^{\{d'\}}~,  \label{eq:cu5a}\\
  \int dV_R \widehat{\Gamma}_{(u)0}^{\{d'\}}
  (\dslash{R}+\mb_h) &\seq
   \Omega_{(u_f)00}^{\{d'\}} +
   \Omega_{(u_f)01}^{\{d'\}} \ptirac~,
   &
  \int dV_R \widehat{\Gamma}_{(u)1}^{\{d'\}} \ptirac 
  (\dslash{R}+\mb_h) &\seq
   \Omega_{(u_f)10}^{\{d'\}} +
   \Omega_{(u_f)11}^{\{d'\}} \ptirac~. \label{eq:cu5b}
  \end{align}
\end{subequations}
The integrals in (\ref{eq:cu5b}) can be evaluated in terms of the
bubble and triangle integrals of \ref{sec:integrals}.  Terms
proportional to $\qirac\,'$ and $\ptirac\qirac\,'$, coming from
$\dslash{C}_1^{(u_f)}$ and $\ptirac \dslash{C}_1^{(u_f)}$, reduce to
the form (\ref{eq:cu5b}) when sandwiched between $\overline{u}'$ and
$u$, since $\qirac\,'\seq \ptirac - \mb_{a'}$ and
$\ptirac\qirac\,'\seq \mb_{a'}\ptirac - \mb_{a'}^2 + \mm_{b'}^2$.
This yields the result,
\begin{equation}
  \label{eq:cu7}  
\begin{aligned}
f^4 \Omega_{(u_f)00}^{\{d'\}} & = \mb_{d'} \mb_h B_0^{hh'} - (\mb_h +
\mb_{a'} -\mb_{d'}) (\mb_h + \mb_{d'})(\mb_{a'} + \mb_{d'}) \left(
  \mb_h C_0^{(u_f)} - \mb_{a'} F_2^{(u_f)}\right), \\
f^4 \Omega_{(u_f)01}^{\{d'\}} & = \mb_{d'} B_1^{hh'} - (\mb_h +
\mb_{a'} -\mb_{d'}) (\mb_h + \mb_{d'})(\mb_{a'} + \mb_{d'}) \left(
  F_1^{(u_f)} + F_2^{(u_f)}\right),  \\
f^4 \Omega_{(u_f)10}^{\{d'\}} & = s B_1^{hh'} + (\mb_h +
\mb_{d'})(\mb_{a'} + \mb_{d'}) \left( s F_1^{(u_f)} - (\mb_{a'}^2 -
  \mm_{b'}^2) F_2^{(u_f)}\right),  \\
f^4 \Omega_{(u_f)11}^{\{d'\}} & = \mb_h B_0^{hh'} + (\mb_h +
\mb_{d'})(\mb_{a'} + \mb_{d'}) \left( \mb_{a'} F_2^{(u_f)} + \mb_{h}
  C_0^{(u_f)}\right). 
\end{aligned}
\end{equation}
The contribution of diagrams $cu$, $su$ to the absorptive part of the
amplitude, given by the terms in (\ref{eq:cutting3}) with $\G'=u$ and
$\G=c,s$, are then determined by (\ref{eq:cu3}), (\ref{eq:cu5})
and (\ref{eq:cu7}) in terms of the scalar integrals of
\ref{sec:integrals}.  

Similarly, the absorptive part of diagrams $(uc)$, $(us)$ is given by
those terms in (\ref{eq:cutting3}) with $\G'=c,s$ and $\G=u$.  In this
case we have,
\begin{equation}
  \label{eq:cu8}
  \widehat{\Gamma}_{(u\G')0}^{\{dd'\}} = \Omega_{(u_i)0}^{\{d\}}
  \widehat{\Gamma}_{(\G')0}^{\{d'\}} + s \Omega_{(u_i)1}^{\{d\}}
  \widehat{\Gamma}_{(\G')1}^{\{d'\}}~,
  \qquad
  \widehat{\Gamma}_{(u\G')1}^{\{dd'\}} = \Omega_{(u_i)0}^{\{d\}}
  \widehat{\Gamma}_{(\G')1}^{\{d'\}} + \Omega_{(u_i)1}^{\{d\}}
  \widehat{\Gamma}_{(\G')0}^{\{d'\}}~. 
\end{equation} 
The form factors $\Omega_{(u_i)0,1}^{\{d\}}$, defined by the analog
of (\ref{eq:cu4})
\begin{equation}
  \label{eq:cu9}
    \int dV_R (\dslash{R}+\mb_h)
    \left(\widehat{\Gamma}_{(u)0}^{\{d\}} + 
    \widehat{\Gamma}_{(u)1}^{\{d\}} \ptirac \right)
   \seq
   \Omega_{(u_i)0}^{\{d\}} +
  \Omega_{(u_i)1}^{\{d\}} \ptirac~,
\end{equation}
are given by (\ref{eq:cu5a}) and (\ref{eq:cu7}) with the substitutions
$(u_f)\rightarrow (u_i)$, $a'\rightarrow a$, $b'\rightarrow b$,
$d'\rightarrow d$ in terms of the scalar integrals $C_0^{(u_i)}$,
$F_1^{(u_i)}$, $F_2^{(u_i)}$ given in \ref{sec:integrals}.

\subsection{Box diagram}
\label{sec:box}

The term in (\ref{eq:cutting3}) with $\G = u = \G'$ is of the form,
\begin{equation}
  \label{eq:uu1}
  \widehat{\Gamma}_{(uu)k}^{\{dd'\}} = \sum_{i,j=0}^1
  \Omega_{(ij)k}^{\{dd'\}}~,
  \qquad
  k=0,1~,
\end{equation}
where $\Omega_{(ij)0,1}^{\{dd'\}}$, $i,j=0,1$, are defined by the
relations
\begin{equation}
  \label{eq:uu2}
  \begin{aligned}
  \int dV_R \widehat{\Gamma}_{(u)0}^{\{d'\}}
  \widehat{\Gamma}_{(u)0}^{\{d\}} (\dslash{R}+\mb_h) 
    & \seq \Omega_{(00)0}^{\{dd'\}} +
  \Omega_{(00)1}^{\{dd'\}} \ptirac~,\\
  \int dV_R \widehat{\Gamma}_{(u)0}^{\{d'\}}
  \widehat{\Gamma}_{(u)1}^{\{d\}} (\dslash{R}+\mb_h) 
     \ptirac& \seq \Omega_{(01)0}^{\{dd'\}} +
  \Omega_{(01)1}^{\{dd'\}} \ptirac~,\\  
  \int dV_R \widehat{\Gamma}_{(u)1}^{\{d'\}}
  \widehat{\Gamma}_{(u)0}^{\{d\}} \ptirac
  (\dslash{R}+\mb_h) & \seq \Omega_{(10)0}^{\{dd'\}} +
  \Omega_{(10)1}^{\{dd'\}} \ptirac~,\\
  \int dV_R \widehat{\Gamma}_{(u)1}^{\{d'\}}
  \widehat{\Gamma}_{(u)1}^{\{d\}} \ptirac
  (\dslash{R}+\mb_h)  \ptirac& \seq
  \Omega_{(11)0}^{\{dd'\}} + 
  \Omega_{(11)1}^{\{dd'\}} \ptirac~.
  \end{aligned}
\end{equation}
Direct evaluation of these integrals, and use of the Dirac equation in
the form $\qirac\ptirac \seq \mb_a \ptirac - \mb_a^2 + \mm_b^2$, $\ptirac
\qirac\,'\ptirac \seq \mb_{a'} s - (\mb_{a'}^2 - \mm_{b'}^2) \ptirac$,
etc., leads to,
  \begin{align}
    f^4 \Omega_{(00)0}^{\{dd'\}} & =  \mb_d \mb_{d'} \mb_h
    B_0^{\{hh'\}} + \mb_{d'} (\mb_a + \mb_h - \mb_{d}) (\mb_h +
    \mb_{d}) (\mb_a + \mb_d) \left[ \mb_a F_2^{(u_i)} - \mb_h
      C_0^{(u_i)} \right]\nonumber\\
    & + \mb_{d} (\mb_{a'} + \mb_h - \mb_{d'}) (\mb_h +
    \mb_{d'}) (\mb_{a'} + \mb_{d'}) \left[ \mb_{a'} F_2^{(u_f)} - \mb_h
      C_0^{(u_f)} \right]\nonumber\\
    & + (\mb_{a'} + \mb_h - \mb_{d'})(\mb_h + \mb_{d'})
  (\mb_{a'} + \mb_{d'}) (\mb_a + \mb_h - \mb_{d}) (\mb_h +
  \mb_{d}) (\mb_a + \mb_d) \nonumber\\
  & \times \left[-\mb_a F_2^{(uu)} -\mb_{a'} F_3^{(uu)} + \mb_h D_0
  \right],\nonumber\\
    f^4 \Omega_{(00)1}^{\{dd'\}} & =  \mb_d \mb_{d'} B_1^{\{hh'\}} -
    \mb_{d'} (\mb_a + \mb_h - \mb_{d}) (\mb_h + \mb_{d}) (\mb_a +
    \mb_d) \left[ F_1^{(u_i)} + F_2^{(u_i)} \right]\nonumber\\
    & - \mb_{d} (\mb_{a'} + \mb_h - \mb_{d'}) (\mb_h +
    \mb_{d'}) (\mb_{a'} + \mb_{d'}) \left[ F_1^{(u_f)} + F_2^{(u_f)}
    \right]\nonumber\\ 
    & + (\mb_{a'} + \mb_h - \mb_{d'})(\mb_h + \mb_{d'})
  (\mb_{a'} + \mb_{d'}) (\mb_a + \mb_h - \mb_{d}) (\mb_h +
  \mb_{d}) (\mb_a + \mb_d) \nonumber\\
  & \times \left[ F_1^{(uu)} + F_2^{(uu)} + F_3^{(uu)} \right],\nonumber\\
    f^4 \Omega_{(01)0}^{\{dd'\}} & =  \mb_{d'} s B_1^{\{hh'\}} + 
    \mb_{d'} (\mb_h + \mb_{d}) (\mb_a + \mb_d) \left[ s F_1^{(u_i)} -
      (\mb_a^2 - \mm_b^2 )F_2^{(u_i)} \right]\nonumber\\
    & - s (\mb_{a'} + \mb_h - \mb_{d'}) (\mb_h + \mb_{d'}) (\mb_{a'} +
    \mb_{d'}) \left[ F_1^{(u_f)} + F_2^{(u_f)} \right]\nonumber\\
    & - (\mb_{a'} + \mb_h - \mb_{d'})(\mb_h + \mb_{d'})
  (\mb_{a'} + \mb_{d'}) (\mb_h + \mb_{d}) (\mb_a + \mb_d) \nonumber\\
  & \times \left[s F_1^{(uu)} - (\mb_{a}^2 -\mm_b^2) F_2^{(uu)} + s
    F_3^{(uu)} \right],\nonumber\\
    f^4 \Omega_{(01)1}^{\{dd'\}} & =  \mb_{d'} \mb_{h} B_0^{\{hh'\}} + 
    \mb_{d'} (\mb_h + \mb_{d}) (\mb_a + \mb_d) \left[ \mb_h C_0^{(u_i)}
    + \mb_a F_2^{(u_i)} \right]\nonumber\\
    & - (\mb_{a'} + \mb_h - \mb_{d'}) (\mb_h + \mb_{d'}) (\mb_{a'} +
    \mb_{d'}) \left[ \mb_h C_0^{(u_f)} - \mb_{a'} F_2^{(u_f)} \right]\nonumber\\
    & - (\mb_{a'} + \mb_h - \mb_{d'})(\mb_h + \mb_{d'})
  (\mb_{a'} + \mb_{d'}) (\mb_h + \mb_{d}) (\mb_a + \mb_d) \nonumber\\
  & \times \left[ \mb_h D_0 + \mb_{a} F_2^{(uu)} - \mb_{a'} F_3^{(uu)}
  \right],\nonumber\\ 
    f^4 \Omega_{(10)0}^{\{dd'\}} & =  \mb_{d} s B_1^{\{hh'\}} + 
    \mb_{d} (\mb_h + \mb_{d'}) (\mb_{a'} + \mb_{d'}) \left[ s F_1^{(u_f)} -
      (\mb_{a'}^2 - \mm_{b'}^2 )F_2^{(u_f)} \right]\nonumber\\
    & - s (\mb_{a} + \mb_h - \mb_{d}) (\mb_h + \mb_{d}) (\mb_{a} +
    \mb_{d}) \left[ F_1^{(u_i)} + F_2^{(u_i)} \right]\nonumber\\
    & - (\mb_{a} + \mb_h - \mb_{d})(\mb_h + \mb_{d'})
  (\mb_{a'} + \mb_{d'}) (\mb_h + \mb_{d}) (\mb_a + \mb_d) \nonumber\\
  & \times \left[s F_1^{(uu)} + s F_2^{(uu)} - (\mb_{a'}^2 -\mm_{b'}^2) 
    F_3^{(uu)} \right],\nonumber\\
    f^4 \Omega_{(10)1}^{\{dd'\}} & =  \mb_{d} \mb_{h} B_0^{\{hh'\}} + 
    \mb_{d} (\mb_h + \mb_{d'}) (\mb_{a'} + \mb_{d'}) \left[ \mb_h C_0^{(u_f)}
    + \mb_{a'} F_2^{(u_f)} \right]\nonumber\\
    & - (\mb_{a} + \mb_h - \mb_{d}) (\mb_h + \mb_{d}) (\mb_{a} +
    \mb_{d}) \left[ \mb_h C_0^{(u_i)} - \mb_{a} F_2^{(u_i)} \right]\nonumber\\
    & - (\mb_{a} + \mb_h - \mb_{d})(\mb_h + \mb_{d'})
  (\mb_{a'} + \mb_{d'}) (\mb_h + \mb_{d}) (\mb_a + \mb_d) \nonumber\\
  & \times \left[ \mb_h D_0 - \mb_{a} F_2^{(uu)} + \mb_{a'} F_3^{(uu)}
  \right],\nonumber\\ 
  f^4 \Omega_{(11)0}^{\{dd'\}} & =  s \mb_{h} B_0^{\{hh'\}} + 
    (\mb_h + \mb_{d}) (\mb_a + \mb_d) s \left[ \mb_h C_0^{(u_i)}
    + \mb_a F_2^{(u_i)} \right]\nonumber\\
    & + (\mb_h + \mb_{d'}) (\mb_{a'} + \mb_{d'}) s \left[ \mb_h
      C_0^{(u_f)} + \mb_{a'} F_2^{(u_f)} \right]\nonumber\\ 
    & + (\mb_h + \mb_{d'})
  (\mb_{a'} + \mb_{d'}) (\mb_h + \mb_{d}) (\mb_a + \mb_d) s%\nonumber\\ 
  \left[ \mb_h D_0 + \mb_{a} F_2^{(uu)} + \mb_{a'} F_3^{(uu)}
  \right],\nonumber\\  
   f^4 \Omega_{(11)1}^{\{dd'\}} & =  s B_1^{\{hh'\}} + 
    (\mb_h + \mb_{d}) (\mb_a + \mb_d) \left[ s F_1^{(u_i)} -
      (\mb_a^2 - \mm_b^2 )F_2^{(u_i)} \right]\nonumber\\
    & + (\mb_h + \mb_{d'}) (\mb_{a'} + \mb_{d'}) \left[ s F_1^{(u_f)} -
      (\mb_{a'}^2 - \mm_{b'}^2 )F_2^{(u_f)} \right]\nonumber\\ 
    & + (\mb_h + \mb_{d'}) (\mb_{a'} + \mb_{d'}) (\mb_h + \mb_{d})
    (\mb_a + \mb_d) \nonumber\\ 
    & \times \left[s F_1^{(uu)} - (\mb_{a}^2 -\mm_b^2) F_2^{(uu)}
     - (\mb_{a'}^2 -\mm_{b'}^2) F_3^{(uu)} \right]. \label{eq:uu3}
  \end{align}
Thus, the contribution to (\ref{eq:cutting3}) from the box diagram
$(uu)$ in fig.\ \ref{fig:fig1} is given by (\ref{eq:uu1}) and
(\ref{eq:uu3}) in terms of the scalar integrals of \ref{sec:integrals}. 

\section{Discussion: numerical results}
\label{sec:discu}

The expression (\ref{eq:polariz}) for the polarization involves the
tree-level scattering amplitude (\ref{eq:tree}), and the absorptive
part of the one-loop amplitude as given explicitly in analytical form
in sect.\ \ref{sec:absptv} and \ref{sec:integrals}. In this
section we discuss those results from a numerical point of view.

For numerical computations we use physical meson and baryon masses,
and coupling constants $D=0.80\pm0.01$ and $F=0.46\pm 0.01$
\cite{clo93} (see also \cite{bor99,rat99}).  The meson weak-decay
constant $f$ should be given, in principle, its chiral-limit value
$f_0 < f_\pi$ which in SU(3) chiral perturbation theory is not
precisely known (see \cite{oll01} and refs.\ cited there).  Since we
are working at leading order we can, alternatively, set $f$ to an
average of its physical values $\langle f_{\pi,K,\eta}\rangle >f_\pi$.
Both possibilities have been used in phenomenological analyses in the
framework of Unitarized Baryon Chiral Perturbation Theory (UBChPT).
Data on meson--baryon scattering cross sections and threshold
branching fractions have been succesfully described in UBChPT with
$f\cong$ 74---86 MeV \cite{oll01}, and with $f = 1.123 f_\pi \cong
103$ MeV \cite{ose98,jid02,ben02}.  In numerical computations we adopt
the latter value, and discuss the dependence on $f$ of our results
below.

If we restrict ourselves strictly to leading-order perturbation theory
the denominator in (\ref{eq:polariz}) must be $|\T_\mathrm{tree}|^2$.
The polarization $\P'_{*\mathrm{l.o.}}$ computed in this way satisfies
$|\P'_{*\mathrm{l.o.}}| \leq 1$ only perturbatively.  That inequality
can be violated if the one-loop absorptive amplitude entering the
asymmetry $\A'_*$ in the numerator of (\ref{eq:polariz}) is not much
smaller than the tree-level amplitude $\T_\mathrm{tree}$, thus
signalling a breakdown of the l.o.\ approximation.  We denote $\P'_*$
the polarization computed using the same amplitude $\T =
\T_\mathrm{tree} + i \T_\mathrm{1-loop, abs.}$ in the denominator of
(\ref{eq:polariz}) as in the numerator, so that $|\P'_*| \leq 1$ holds
exactly.  We may consider the difference of $\P'_*$ and
$\P'_{*\mathrm{l.o.}}$ as a rough measure of the validity of the l.o.\
approximation for the polarization.  From (\ref{eq:tree}) and fig.\
\ref{fig:fig1} we find that $\T_\mathrm{tree} \propto 1/f^2$,
$\T_\mathrm{1-loop,abs.} \propto 1/f^4$, and $\A \propto 1/f^6$.
Therefore, $\P'_{*\mathrm{l.o.}}  \propto 1/f^2$ and $\P'_* \sim
1/(c_1 f^2 + c_2 + c_3/f^2)$, with $c_{1,2,3}$ some $f$-independent
coefficients.  Thus, the difference $|\P'_* - \P'_{*\mathrm{l.o.}}|$
is expected to increase with decreasing $f$.  In those processes and
kinematic regions where the l.o.\ approximation is accurate we expect
both $\P'_{*\mathrm{l.o.}}$ and $\P'_*$ to scale with $f$ as $1/f^2$,
the former exactly and the latter approximately.

In figs.\ \ref{fig:fig2}---\ref{fig:fig5} we plot the
polarization $\P'_*$ as a function of the center-of-mass scattering
angle for several reactions and values of $\sqrt{s}$, as an
illustration of the results obtained in the previous sections directly
from BChPT.  $\P'_{*\mathrm{l.o.}}$ is also shown in those figures for
comparison. 

For $p\pi^\pm\rightarrow p\pi^\pm$ and $p\pi^-\rightarrow n\pi^0$,
curves (1)--(3) in fig.\ \ref{fig:fig2} correspond to energies more
than 100 MeV below the $\Delta$-resonance peak, in the region where
BChPT should be applicable.  $\P'_*$ and $\P'_{*\mathrm{l.o.}}$ do not
differ appreciably at those energies, except for curve (3) for
$p\pi^-\rightarrow p\pi^-$, pointing to a good convergence of
perturbation theory in that energy region.  As seen in the figure,
$\P'_*$ reaches sizable values for the elastic processes.  For
$q_\mathrm{lab}\gtrsim 100$ MeV not only higher-order corrections are
expected to become important, but also the $\Delta$ resonance
contributions are essential.  In fig.\ \ref{fig:fig3} we plot the
polarization $\P'_*$ including the contribution from the pole in
$s$-channel $\Delta$ resonance exchange.  The latter, not contained in
our analytical results, was computed numerically from the scattering
amplitudes in \cite{mei00}.  As seen from the figure, the $\Delta$
resonance contribution to $\P'_*$ is already quantitatively
significant at $q_\mathrm{lab} = 125$ MeV, becoming the dominant one
at higher energies.  As expected, within the resonance peak our
analytical results are not applicable.

Our results for $pK^+\rightarrow pK^+$ are shown in fig.\
\ref{fig:fig4}.  For this process we expect BChPT to be valid over the
entire energy range of the figure, whose highest $\sqrt{s}$ is about
250 MeV below the $\Delta^{++}K^0$/$\Delta^{+}K^+$ rest mass at the
$\Delta$ peak.  The difference of $\P'_*$ and $\P'_{*\mathrm{l.o.}}$
is large for curve (6) in fig.\ \ref{fig:fig4}, reflecting the fact
that higher-order corrections become important at about
$q_\mathrm{lab}\sim 300$ MeV.  As seen in the figures, the
polarizations $\P'_*$ in $p\pi^+\rightarrow p\pi^+$ and
$pK^+\rightarrow pK^+$ are in relation $(\P'_*)_{p\pi^+} \sim 10
(\P'_*)_{pK^+}$ at fixed initial meson momentum and
$-1<\cos\theta_\mathrm{cm}<1$. This fact deserves some consideration
since we expect the dynamics to be similar in both elastic processes,
except for the stronger coupling in the $S=1$ channel due to the
larger kaon mass.\footnote{Notice that for both $p\pi$ interactions in
  the $I=3/2$ sector and $pK$ with $I=1$ the scattering lengths are
  negative, indicating a repulsive interaction \cite{ber95}.}
Restricting ourselves to $0\leq q_\mathrm{lab}\leq 125$ MeV where the
difference between $|\T|^2$ and $|\T_\mathrm{tree}|^2$ is negligible,
for the spin asymmetry $\A'_*$ defined in (\ref{eq:asym}) and
(\ref{eq:spina}) numerically we find, roughly, $(\A'_*)_{p\pi^+} \sim
1/10 (\A'_*)_{pK^+}$.  The factor $|\vec{q}\wedge \vec{p}|$ in
(\ref{eq:spina}) is almost equal in both processes.  Whereas in
$p\pi^+$ elastic scattering the two terms in the factor $( \Re\Gamma_0
\Im\Gamma_1 - \Im\Gamma_0 \Re\Gamma_1)$ in (\ref{eq:asym}) partially
cancel, in $pK^+$ they have the same sign, leading to a larger spin
asymmetry for the latter process.\footnote{The crucial sign difference
  between the two elastic processes comes from $\Im\Gamma_1$.}  From
(\ref{eq:totsqam}), $|\T|^2$ is a sum of three terms $C\Gamma^2$
which, for each of the two processes, are all of the same order of
magnitude.  Numerically, we have $(C\Gamma^2)_{p\pi^+} \sim 10
(C\Gamma^2)_{pK^+}$ \footnote{The factors $C_{ij}$ in
  (\ref{eq:totsqam}) are essentially equal for the two elastic
  processes.  The factor $\sim 10$ referred to here comes from the
  squared tree-level amplitudes. Specifically, the $u$-channel
  contribution in $pK^+$ scattering is smaller than in $p\pi^+$, and
  that of the $c$-channel is almost equal for both processes.}  which
suggests an analogous relation holds for $|\T|^2$, therefore
apparently leading to the wrong result $(\P'_*)_{p\pi^+} \sim 1/100
(\P'_*)_{pK^+}$. In fact, due to destructive interference between the
spin-flip and non-spin-flip terms in the tree-level amplitude, we have
$|\T|^2_{pK^+} \sim 1/10 (C\Gamma^2)_{pK^+}$ and a much stronger
effect in $p\pi^+$, $|\T|^2_{p\pi^+} \sim 1/10^4
(C\Gamma^2)_{p\pi^+}$.  Therefore, we get $|\T|^2_{p\pi^+} \sim 1/10^2
|\T|^2_{pK^+}$ which, together with the above relation for $\A'_*$,
results in $(\P'_*)_{p\pi^+} \sim 10 (\P'_*)_{pK^+}$.

Due to the strong coupling in the $S=-1$ sector
\cite{ber07a,oll01,ose98,lee94} BChPT is not directly applicable to
nucleon--antikaon processes.  Rather, higher-order corrections must be
resummed with unitarization techniques such as UBChPT.  In the $I=0$
channels the $\Lambda(1405)$ resonance lies $\sim$25 MeV below
threshold, dominating the dynamics of $pK^- \rightarrow \Sigma\pi$
near threshold in the $S$ wave.  In the $I=1$ channels
$pK^-\rightarrow \Lambda\pi^0, \Sigma\pi$ the narrow decuplet
$\Sigma(1385)$ lies $\sim$46 MeV below threshold.  The effect on
final-state polarization of such strong-coupling phenomena, not taken
into account in this paper, will be discussed elsewhere.  For
completitude, however, we illustrate our BChPT results for $\P'_*$ for
several $p K^-$ scattering processes in fig.\ \ref{fig:fig5}, about
which we shall make some qualitative remarks.  As seen there, the
difference of $\P'_*$ and $\P'_{*\mathrm{l.o.}}$ is relatively small
in the $\Lambda\pi^0$ channel but very large in the $\Sigma^-\pi^+$
channel, where $|\P'_{*\mathrm{l.o.}}|>1$ over most of the range of
$\cos \theta_\mathrm{cm}$, whereas the other two channels are
intermediate between those cases, with large $|\P'_* -
\P'_{*\mathrm{l.o.}}|$ but $|\P'_{*\mathrm{l.o.}}|<1$.  Such large
contributions from the one-loop absorptive part signal the
inapplicability of perturbation theory.  The sign of $\P'_*$ in the
$\Sigma^-\pi^+$ channel is opposite to that of the other channels, and
the maximum of $|\P'_*|$ over the range $-1\leq \cos
\theta_\mathrm{cm} \leq 1$ shows a much weaker dependence on energy in
that channel than in the others.

\section{Final Remarks}
\label{sec:final}

In this paper we computed the spin asymmetry and the polarization for
the final-state baryon in unpolarized two-body meson--baryon
scattering in lowest non-trivial order BChPT.  The spin asymmetry
(\ref{eq:asym}) is a purely quantum effect which arises from the
interference of the spin-flip and non-spin-flip amplitudes when
on-shell intermediate channels are open.  More precisely, the
asymmetry is proportional to $\Im(\Gamma_0\Gamma_1^*)$, which must
vanish if the imaginary parts of both factors are zero as happens at
tree level.  The expression (\ref{eq:asym}) for $\A$ is a direct
consequence of the form (\ref{eq:parameta}) of the amplitude, which in
turn follows from Lorentz covariance, the discrete spacetime
symmetries of the strong interactions, and the spin and parity quantum
numbers of ground-state mesons and baryons.

The scattering amplitudes are computed above in a physical flavor
basis, incorporating baryon mass splittings already in the tree-level
expressions (\ref{eq:tree}).  As mentioned in Sect.\ \ref{sec:lag},
including higher-order flavor-breaking effects is not inconsistent at
leading order. The necessary absorptive parts are obtained at one-loop
level from the purely dispersive tree-level amplitudes by means of
unitarity relations.  Those are given in Sect.\ \ref{sec:absptv} and
in \ref{sec:integrals} in closed analytical form, a result which
is interesting by itself due to its applicability in other
perturbative computations.

A numerical analysis of our results for the spin asymmetry and
polarization is carried out in Sect.\ \ref{sec:discu}, where several
meson--proton processes are considered.  Polarization effects are seen
there to become stronger the higher the energy of the process.  Yet,
even at the very low energies above threshold at which BChPT can
reasonably be expected to be applicable, sizable values of
polarization are found in some processes, both in elastic and
inelastic reaction channels.  In elastic $N\pi$ scattering
polarizations in the range $\sim$ 10---25\% are found.  By contrast,
for the reasons analyzed above, polarizations for $pK^+\rightarrow
pK^+$ are smaller.  In the $S=-1$ meson--baryon sector the
leading-order approximation used here cannot be expected to be valid.
In fact, it is known that due to the strong coupling and subthreshold
resonances ($\Lambda$(1405)) in this sector, non-perturbative
coupled-channels analysis are required to reproduce available
cross-section data.  The polarization curves in fig.\ \ref{fig:fig5}
are only meant to illustrate the leading-order results obtained here
for the polarization.

As discussed in sect.\ \ref{sec:discu}, a source of uncertainty in the
leading-order result is the value of $f$.  That uncertainty is
inherent in the leading-order approximation, since it is higher-order
corrections that shift $f$ from its chiral-limit value and split it
into its physical values.  On the other hand, the chiral-limit $f$ is
subject to considerable uncertainty itself in the three-flavor case.
The effects of a variation in $f$ on our results are readily
quantifiable, however, since we expect the polarization to scale
approximately as $1/f^2$ (and exactly so at leading order).  We remark
that the numerical value for $f$ used in sect.\ \ref{sec:discu} is
conservative in this respect, with lower values of $f$ resulting in
polarizations larger (up to a factor of about 2 for $f$ as low as 76
MeV) than those reported in figs.\ \ref{fig:fig2}---\ref{fig:fig5}.

Finally, we hope that the results obtained here may prompt a
re-analysis of the wealth of data obtained in many experiments with
meson beams ($\pi$ beams at LAMPF, TRIUMF and PSI, for instance) in
the past decades, leading to experimental information on the
observables discussed here.

\section*{Acknowledgements}

I would like to thank Prof.\ C.\ A.\ Garc\'{\i}a Canal for his useful
comments on a preliminary version of this paper.

\appendix

\section{Kinematics}
\label{sec:loops}

In this appendix we gather some kinematical definitions used 
throughout the paper.  We introduce the notation
\begin{equation}
  \label{eq:omega}
  \omega(x,y,z) = (x^2+y^2+z^2-2xy-2xz-2yz)^\frac{1}{2}
  = (x-(\sqrt{y}+\sqrt{z})^2)^\frac{1}{2}
  (x-(\sqrt{y}-\sqrt{z})^2)^\frac{1}{2}~. 
\end{equation}
The function $\omega$ appears frequently in relativistic kinematics
(\emph{e.g.,} in the center of mass frame $|\vec{p}| =
\omega(s,\mb_a^2, \mm_{b}^2)/(2\sqrt{s})$). The Mandelstam invariants
for the process $| B^{a}(p,\sigma) M^b(q)\rangle \longrightarrow |
B^{a'}(p',\sigma') M^{b'}(q')\rangle$ are
\begin{equation}
  \label{eq:mandelstam}
  s=(p+q)^2=(p'+q')^2~,
  \quad
  t=(p-p')^2=(q-q')^2~,
  \quad
  u=(p-q')^2=(p'-q)^2~,
\end{equation}
with $s+t+u = \mb_a^2 + \mb_{a'}^2 + \mm_{b}^2 + \mm_{b'}^2~$.
The physical region for the process is defined by the inequalities
\begin{equation}
  \label{eq:physreg}
  s_\sss{\mathrm{th}}\leq s~,
  \quad
  \tmin \leq t \leq \tmax~,  
  \quad
  \umin \leq u \leq \umax~,
\end{equation}
where,
\begin{equation}
  \label{eq:kinlim}
  \begin{aligned}
    s_\sss{\mathrm{th}} &= \max\left\{(\mb_a + \mm_b)^2, (\mb_{a'} +
      \mm_{b'})^2 \right\}\\
    t_{\substack{\sss{\mathrm{max}}\\[-2pt]\sss{\mathrm{min}}}} &=
    -\frac{1}{2s} \left( \rule{0pt}{10pt} s^2 - s (\mb_a^2 +
      \mb_{a'}^2 + \mm_b^2 + \mm_{b'}^2 ) + (\mb_a^2 - \mm_b^2)
      (\mb_{a'}^2 - \mm_{b'}^2) \right) \\
    &\quad \pm \frac{1}{2s} \omega(s,\mb_a^2,\mm_b^2)
      \omega(s,\mb_{a'}^2,\mm_{b'}^2),\\
    u_{\substack{\sss{\mathrm{max}}\\[-2pt]\sss{\mathrm{min}}}} &=
    -\frac{1}{2s} \left( \rule{0pt}{10pt} s^2 - s (\mb_a^2 +
      \mb_{a'}^2 + \mm_b^2 + \mm_{b'}^2 ) - (\mb_a^2 - \mm_b^2)
      (\mb_{a'}^2 - \mm_{b'}^2) \right) \\
    &\quad \pm \frac{1}{2s}
    \omega(s,\mb_a^2,\mm_b^2) \omega(s,\mb_{a'}^2,\mm_{b'}^2).\\    
  \end{aligned}
\end{equation}

\setcounter{equation}{0}
\section{Phase-space integrals}
\label{sec:integrals}

In this section we collect analytical results for some phase-space
integrals.  The notation we use is analogous to that for the loop
integrals of which they are absorptive (or, for scalar integrals,
imaginary) parts.  The analytical expressions, obtained by standard
methods \cite{tho79,den91,pas79}, hold only in the physical region
defined in (\ref{eq:physreg}).  The volume element is in all cases
\begin{equation}
  \label{eq:volume}
dV_R \equiv d^4\!R\, \delta_+(R^2 - \mb_{h}^2) \delta_+((\pt-R)^2 - \mm_{h'}^2),
\end{equation}
where $\delta_+(x^2) \equiv \delta(x^2) \theta(x^0)$ for any
(timelike) four-vector $x^\mu$, with $\theta$ a unit step function.

\subsection{Bubble diagrams}
\begin{equation}
\label{eq:bubble1}
\begin{aligned}
  B_0^{\{hh'\}} &= \int dV_R\, 1 = \frac{\pi}{2}
  \frac{\omega(s,\mb_h^2,\mm_{h'}^2)}{s} \theta(\sqrt{s}- \mb_h 
  -\mm_{h'}) \\
  B_1^{\{hh'\}} &= \frac{1}{s} \int dV_R\, \pt\cdot R  =
  \frac{s+\mb_h^2-\mm_{h'}^2}{2s} B_0^{\{hh'\}} \\ 
  B_1^{\{hh'\}\mu} &= \int dV_R\, R^\mu = B_1^{\{hh'\}} \pt^\mu
\end{aligned}
\end{equation}

\subsection{Triangle diagrams}

Integrals related to triangle diagrams are generically denoted by $C$.
To avoid having to attach long lists of arguments to that symbol, we
distinguish between diagrams in which the external particles directly
attached to the trivalent vertices of the triangle are those of the
final or of the initial state.  In the former case integrals are
denoted $C^{(u_f)}$, and in the latter $C^{(u_i)}$ (see fig.\
\ref{fig:ufui}). 
 
\begin{figure}
  \centering
\includegraphics{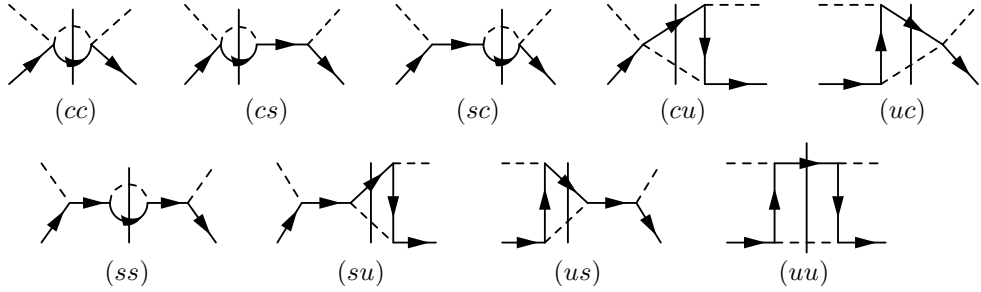}
\caption{Feynman diagrams contributing to the absorptive part of the
  scattering amplitude.  Vertical lines denote unitary cuts factoring
  each diagram into two tree-level diagrams as indicated by the
  labels.}
  \label{fig:fig1}
\end{figure}

\begin{figure}[p]
  \centering
\includegraphics{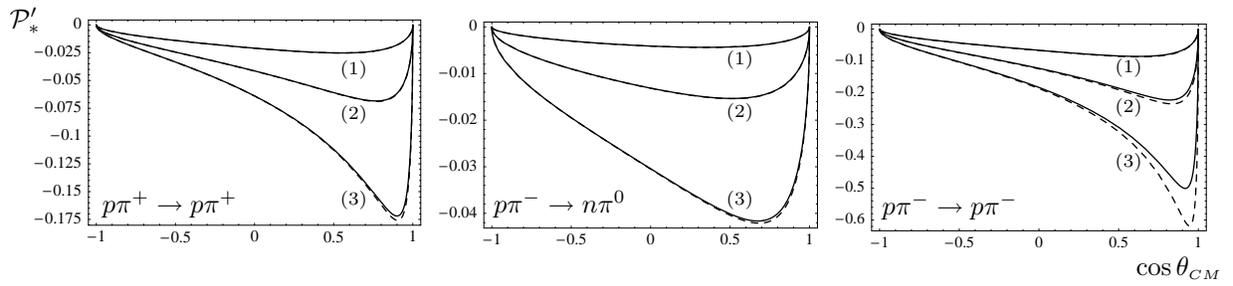}
\caption{BChPT result for the polarization of the final baryon in its
  rest frame as a function of center-of-mass scattering angle.  Solid
  lines: $\P'_*$, dashed lines:$\P'_{*\mathrm{l.o.}}$ (see sect.\
  \ref{sec:discu}).  Curves (1)--(3) correspond to lab frame initial
  meson momentum $q_{\mathrm{lab}} = $ 75, 100, 125 MeV, resp.}
  \label{fig:fig2}
\end{figure}

\begin{figure}[p]
  \centering
\includegraphics{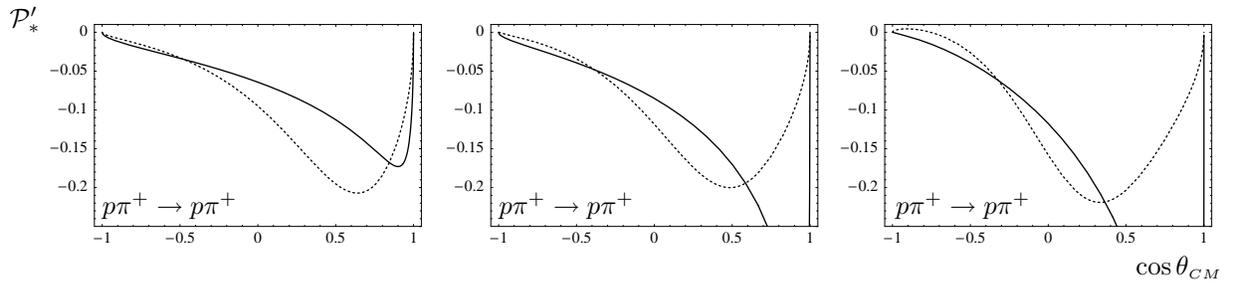}
  \caption{Solid lines as in fig.\ \ref{fig:fig2}, dotted lines:
    $\P'_*$ including $\Delta$ resonance. From left to
    right, $q_{\mathrm{lab}} = $ 125, 150, 200 MeV.}
  \label{fig:fig3}
\end{figure}

\begin{figure}[p]
  \centering
\includegraphics{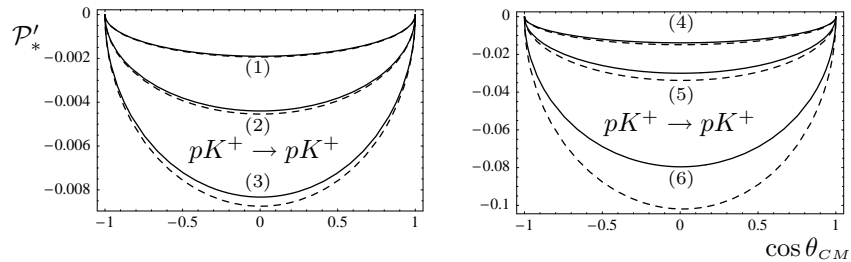}
  \caption{Solid and dashed lines as in fig.\
    \ref{fig:fig2}. Curves (1)--(6): $q_{\mathrm{lab}} = 
    $ 75, 100, 125, 150, 200, 300 MeV, resp.}
  \label{fig:fig4}
\end{figure}

\begin{figure}[p]
  \centering
\includegraphics{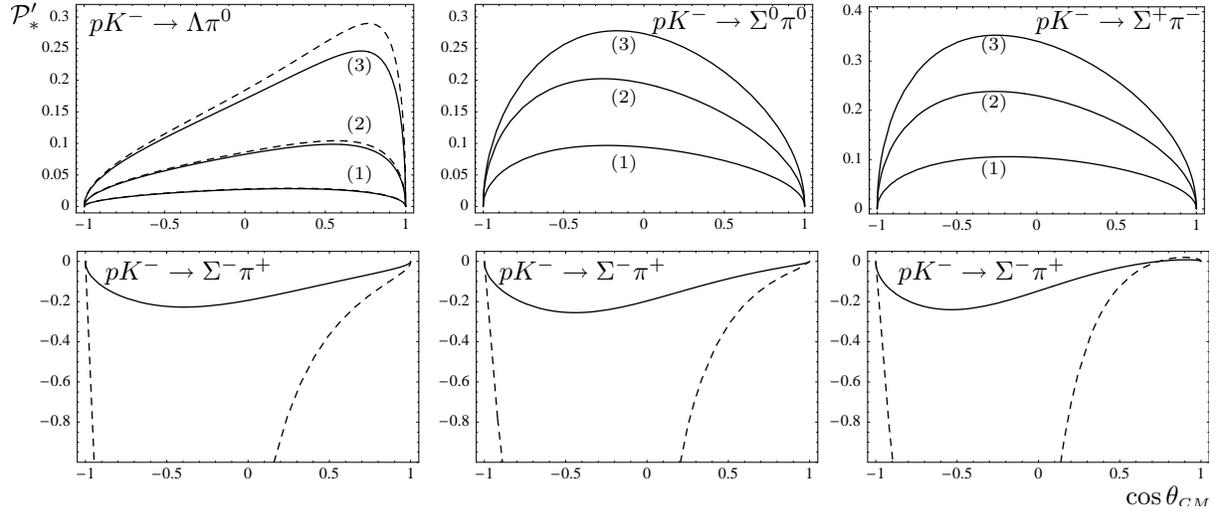}
  \caption{Solid and dashed lines (not shown in some plots for
    clarity) as in fig.\ \ref{fig:fig2}.  Upper row, curves
    (1)--(3): $q_{\mathrm{lab}} = $ 100, 200, 300 MeV, resp. Lower
    row, from left to right: same values of $q_{\mathrm{lab}}$.}
  \label{fig:fig5}
\end{figure}

\begin{figure}
  \centering
  \scalebox{0.75}{
   \includegraphics{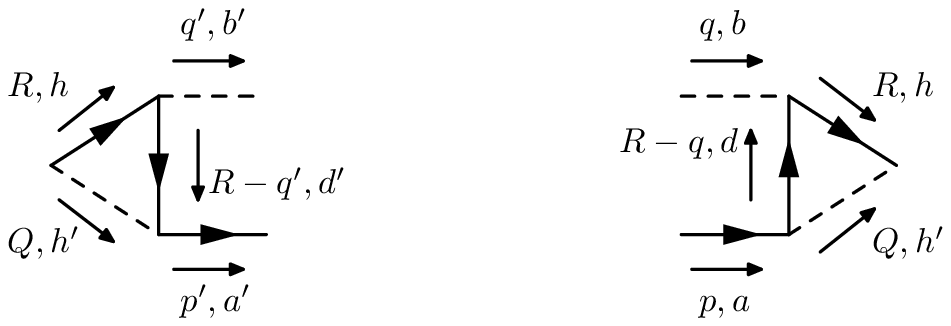}}
\caption{Labelling of momenta and flavors in triangle
  graphs with final- and initial-state particles attached.}
  \label{fig:ufui}
\end{figure}

For integrals of type $C^{(u_f)}$ (diagrams $cu$ and $su$ in fig.\
\ref{fig:fig1}) the integrand depends on $u_f = (R-q')^2$.  The
kinematic limits
$u_{f_{\substack{\sss{\mathrm{max}}\\\sss{\mathrm{min}}}}}$ for
$u_f$ are obtained from
$u_{\substack{\sss{\mathrm{max}}\\[-2pt]\sss{\mathrm{min}}}}$ in
(\ref{eq:kinlim}) with the substitutions $a\rightarrow h$,
$b\rightarrow h'$.  The scalar integrals are,
\begin{align}
  C_0^{(u_f)} &= \int dV_R \frac{1}{u_f -\mb_{d'}^2} =
  -\frac{\pi}{2}
  \frac{\theta(\sqrt{s}-\mb_h-\mm_{h'})}{\omega(s,\mb_{a'}^2,\mm_{b'}^2)}
  \log\left( \frac{\mb_{d'}^2 - \ufmin}{\mb_{d'}^2 -
      \ufmax} \right) \label{eq:triangle1}\\
  C_{1\pt}^{(u_f)} &= \int dV_R \frac{\pt\cdot R}{u_f -\mb_{d'}^2}
  = \frac{s+\mb_h^2 - \mm_{h'}^2}{2} C_0^{(u_f)}
  \label{eq:triangle2}\\ 
  C_{1q'}^{(u_f)} &= \int dV_R \frac{q'\cdot R}{u_f -\mb_{d'}^2}
  = \frac{1}{2} (\mb_h^2 +\mm_{h'}^2 - \mb_{d'}^2) C_0^{(u_f)} -
  \frac{1}{2} B_0^{\{hh'\}} \label{eq:triangle3}
  \end{align}
For the vector integral we introduce two sets of form factors,
$G_{1,2}^{(u_f)}$ are obtained by orthogonal projection and
algebraic reduction, and $F_{1,2}^{(u_f)}$ are given in terms of
$G_{1,2}^{(u_f)}$. 
\begin{equation}
\label{eq:triangle4}
  \begin{gathered}
  C_{1}^{(u_f)\mu} =
    \int dV_R \frac{R^\mu}{u_f -\mb_{d'}^2}
     =  \displaystyle G_{1}^{(u_f)} \ptirac + G_{2}^{(u_f)}
    \left(\qirac\,' - 
      \frac{s-\mb_{a'}^2 + \mm_{b'}^2}{2s} \ptirac\right) 
     =  F_{1}^{(u_f)} \ptirac + F_{2}^{(u_f)} \qirac\,'~,\\ 
     G_{1}^{(u_f)} = \frac{1}{s} C_{1\pt}^{(u_f)}~,
     \quad
  G_{2}^{(u_f)} = -\frac{4 s}{\omega(s,\mb_{a'}^2,\mm_{b'}^2)^2} \left(
    C_{1q'}^{(u_f)} - \frac{s-\mb_{a'}^2 + \mm_{b'}^2}{2 s}
    C_{1\pt}^{(u_f)} \right), \\
  F_{1}^{(u_f)} = G_{1}^{(u_f)} - \frac{s-\mb_{a'}^2 + \mm_{b'}^2}{2
    s} G_{2}^{(u_f)}~,
  \quad
  F_{2}^{(u_f)} = G_{2}^{(u_f)}~.
\end{gathered}
\end{equation}
For integrals of type $C^{(u_i)}$ (diagrams $uc$ and $us$ in fig.\
\ref{fig:fig1}) the integrand depends on $u_i = (R-q)^2$.  The
kinematic limits
$u_{i_{\substack{\sss{\mathrm{max}}\\\sss{\mathrm{min}}}}}$ for $u_i$
are obtained from
$u_{\substack{\sss{\mathrm{max}}\\[-2pt]\sss{\mathrm{min}}}}$ in
(\ref{eq:kinlim}) with the substitutions $a'\rightarrow h$,
$b'\rightarrow h'$.  The scalar integrals $C_0^{(u_i)}$,
$C_{1\pt}^{(u_i)}$ and $C_{1q}^{(u_i)}$, as well as the vector
integral $C_{1}^{(u_i)\mu}$ and its form factors $G_{1,2}^{(u_i)}$,
$F_{1,2}^{(u_i)}$ are obtained from 
(\ref{eq:triangle1})--(\ref{eq:triangle4}) by means of the
substitutions $a'\rightarrow a$, $b'\rightarrow b$, $d'\rightarrow d$,
$p'\rightarrow p$, $q'\rightarrow q$, $u_{f}\rightarrow u_{i}$,
$u_{f_{\substack{\sss{\mathrm{max}}\\\sss{\mathrm{min}}}}}\rightarrow
u_{i_{\substack{\sss{\mathrm{max}}\\\sss{\mathrm{min}}}}}$.

\subsection{Box diagram}

The basic scalar box integral is given by
\begin{equation}
  \label{eq:box1}
\begin{gathered}
  D_0 = \int dV_R \frac{1}{u_f -\mb_{d'}^2} \frac{1}{u_i -\mb_{d}^2}
  = 4\pi \theta(\sqrt{s} - (\mb_h + \mm_{h'}))
  \frac{|\vec{R}|_\mathrm{cm}}{\sqrt{s} \sqrt{\D_{fi}^2 - 4
      \D_f \D_i}}
  \mathrm{Argtanh}\left(\sqrt{\frac{\D_{fi}-2\sqrt{\D_f
          \D_i}}{\D_{fi}+2\sqrt{\D_f \D_i}}}\right),   \\
  |\vec{R}|_\mathrm{cm} =
  \frac{\omega(s,\mb_h^2,\mm_{h'}^2)}{2\sqrt{s}}~,
  \qquad
  \left(\vec{q}\,'\cdot \vec{q}\right)_{\mathrm{cm}} = \frac{1}{2}
  \left( \frac{\umax + \umin}{2} - u \right), \\
  \D_f = (\mb_{d'}^2 - \ufmax) (\mb_{d'}^2 -
  \ufmin),
  \qquad
  \D_i = (\mb_{d}^2 - \uimax) (\mb_{d}^2 -
  u_{i_\mathrm{min}}),\\
  \D_{fi} = \frac{1}{2}( 2 \mb_{d'}^2 - \ufmax -
  \ufmin) ( 2 \mb_{d}^2 - \uimax -
  \uimin) - 8 |\vec{R}|_{\mathrm{cm}}^2
  \left(\vec{q'}\cdot \vec{q}\right)_{\mathrm{cm}}~.  
\end{gathered}
\end{equation}
For all diagrams allowed by conservation laws we must have $\ufmax\leq
\mb_{d'}^2$, $\uimax\leq\mb_{d}^2$ due to the stability of
ground-state baryons against strong decay, therefore $\D_f \geq 0 \leq
\D_i$.  Furthermore, it can be seen from its definition above that
within the physical region $\D_{fi} \geq 2\sqrt{\D_f \D_i}$.  From 
(\ref{eq:box1}), by algebraic reduction we get,
\begin{align}
  D_{1\pt} &= \int dV_R \frac{R\cdot\pt}{(u_f -\mb_{d'}^2)(u_i
    -\mb_{d}^2)} = \frac{1}{2} (s+\mb_h^2 - \mm_{h'}^2) D_0~,
  \label{eq:box2}\\
  D_{1q} &= \int dV_R \frac{R\cdot q}{(u_f -\mb_{d'}^2)(u_i
    -\mb_{d}^2)} = \frac{1}{2} (\mb_h^2 + \mm_{b}^2-\mb_d^2) D_0
  -\frac{1}{2} C_0^{(u_f)}~, \label{eq:box3}\\
  D_{1q'} &= \int dV_R \frac{R\cdot q'}{(u_f -\mb_{d'}^2)(u_i
    -\mb_{d}^2)} = \frac{1}{2} (\mb_h^2 + \mm_{b'}^2-\mb_{d'}^2) D_0 
  -\frac{1}{2} C_0^{(u_i)}~.\label{eq:box4}
\end{align}
As with triangle diagrams, we decompose the vector integral into two
sets of form factors, $G_i^{(uu)}$  and $F_i^{(uu)}$,
\begin{equation}
\label{eq:box5}
  \begin{gathered}
  D_1^\mu = \int dV_R \frac{R^\mu}{(u_f -\mb_{d'}^2)(u_i
    -\mb_{d}^2)}  =  G_1^{(uu)} \ptirac +  G_2^{(uu)} \dslash{v}_2 +
  G_3^{(uu)} \dslash{v}_3 
     = F_1^{(uu)} \ptirac +  F_2^{(uu)} \qirac + F_3^{(uu)}
  \qirac\,'~, \\
  v_2  =  q - \frac{1}{2s}(s+\mm_b^2 - \mb_a^2)\pt~, 
\quad
  v_3  = q' - \frac{1}{2s}(s+\mm_{b'}^{2} -
    \mb_{a'}^2)\pt + \frac{2 s}{\omega(s,\mb_a^2,\mm_b^2)^2} \left(
      u - \frac{\umax + \umin}{2}\right) v_2 ~.
  \end{gathered}
\end{equation}
Calling $v_1=\pt$ we have $v_i\cdot v_j \propto \delta_{ij}$,
$i,j=1,2,3$.  The vectors $v_i$ are obtained from $\pt$, $q$, $q'$ by
Gramm--Schmidt orthogonalization for simplicity.  More symmetrical
results can be obtained with the orthogonalization method of
\cite{sch70}.  The form factors are given by,
\begin{equation}
\label{eq:box6}
\begin{aligned}
  G_1^{(uu)} &= \frac{s+\mb_h^2-\mm_{h'}^2}{2s} D_0~, \\
  G_2^{(uu)} &= \frac{-4 s}{\omega(s,\mb_a^2,\mm_b^2)^2} \left(
    D_{1q} - \frac{1}{2s} (s+\mm_b^2 - \mb_a^2) D_{1\pt}\right),\\
  G_3^{(uu)} &=
  \frac{\omega(s,\mb_a^2,\mm_b^2)^2}{s(t-\tmax)(t-\tmin)} \left( 
    D_{1q'} - \frac{1}{2} (s+\mm_{b'}^2 - \mb_{a'}^2) G_1^{(uu)} -
    \frac{1}{2} \left( u -\frac{\umax + \umin}{2} \right)
    G_2^{(uu)} \right), \\
  F_1^{(uu)} &= G_1^{(uu)} - \frac{s+\mm_b^2-\mb_a^2}{2s} G_2^{(uu)} -
  \frac{s+\mm_{b'}^2 - \mb_{a'}^2}{2s} G_3^{(uu)} -
  \frac{s + \mm_{b}^2 - \mb_{a}^2}{\omega(s,\mb_a^2,\mm_b^2)^2} \left(
    u - \frac{\umax + \umin}{2} \right) G_3^{(uu)}~,\\
  F_2^{(uu)} &= G_2^{(uu)} + \frac{2s}{\omega(s,\mb_a^2,\mm_b^2)^2} 
  \left( u - \frac{\umax + \umin}{2} \right) G_3^{(uu)}~,\\
  F_3^{(uu)} &= G_3^{(uu)}~.
\end{aligned}
\end{equation}
From the definition of $D_1^\mu$ in (\ref{eq:box5}) it is clear that
under exchange of initial and final state ($a\leftrightarrow a'$,
$b\leftrightarrow b'$, $q\leftrightarrow q'$, $p\leftrightarrow p'$,
etc.) we must have $F_1^{(uu)}\leftrightarrow F_1^{(uu)}$ and
$F_2^{(uu)}\leftrightarrow F_3^{(uu)}$.  In the expressions
(\ref{eq:box6}) for $F_{1,2,3}^{(uu)}$ this symmetry is not apparent,
but it has been checked numerically.



\begin{thebibliography}{99}
\bibitem{ber07}V.\ Bernard, U.\ G.\ Meissner, Annu.\ Rev.\ Nucl.\
  Part.\ Sci.\  \textbf{57} (2007) 33.
%
\bibitem{bor07}B.\ Borasoy, \emph{Introduction to Chiral Perturbation 
    Theory,} Lectures given at the 2$^\mathrm{nd}$ Summer School on
  Particle Accelerators and Detectors, Bodrum, Turkey, Sep.\ 2006; 
  arxiv:hep-ph/0703297.
%
\bibitem{sch05}S.\ Scherer, M.\ R.\ Schindler, \emph{A Chiral
    Perturbation Theory Primer,} lectures given at the European Centre
  for Theoretical Studies in Nuclear Physics and Related Areas,
  Trento, Italy, 2005; arxiv:hep-ph/0505265.
%
\bibitem{wei96} S.\ Weinberg, \emph{The Quantum Theory of Fields},
  Vol.\ II, Cambridge Univ.\ Press, New York, 1996.
%
\bibitem{don94}J.\ F.\ Donoghue, E.\ Golowich, B.\ R.\ Holstein,
  \emph{Dynamics of the Standard Model,} Cambridge Univ.\ Press, New 
  York, 1994. 
%
\bibitem{ber07a}V.\ Bernard, \emph{Chiral Perturbation Theory and
    Baryon Properties,} arXiv:0706.0312. 
%
\bibitem{ber95} V.\  Bernard, N.\  Kaiser, U.-G. Meissner, Int.\ J.\
  Mod.\ Phys.\ E \textbf{4} (1995) 193.
%
\bibitem{lea01} E.\ Leader, \emph{Spin in Particle Physics,} Cambridge
  Univ.\ Press, New York, 2001. 
%
\bibitem{sevx} M.\ E.\ Sevior et al., Phys.\ Rev. C \textbf{40} (1989)
  2780. 
%
\bibitem{hofx1} G.\ J.\ Hoffman et al., Phys.\ Rev. C \textbf{58} (1998)
  3484. 
%
\bibitem{hofx2} G.\ J.\ Hoffman et al., Phys.\ Rev. C \textbf{68} (2003)
  018202.
%
\bibitem{wiex} R.\ Wieser et al., Phys.\ Rev. C \textbf{54} (1996)
  1930. 
%
\bibitem{meix} R.\ Meier et al., Phys.\ Lett.\ B \textbf{588} (2004)
  155. 
%
\bibitem{lovx} B.\ R.\ Lovett et al., Phys.\ Rev. D \textbf{23} (1981)
  1924. 
%
\bibitem{fel99} J.\ F\'elix, Mod.\ Phys.\ Lett.\ A \textbf{14} (1999)
  827. 
%
\bibitem{gas88}J.\ Gasser, M.\ E.\ Sainio, A.\ Svark, Nucl.\ Phys.\
  \textbf{B 307} (1988) 779.
%
\bibitem{gas84}J.\ Gasser, H.\ Leutwyler, Ann.\ Phys.\ (N.Y.)
  \textbf{158} (1984) 142.
%
\bibitem{gas85a}J.\ Gasser, H.\ Leutwyler, Nucl.\ Phys.\ B
  \textbf{250} (1985) 465.
%
\bibitem{kra90}A.\ Krause, Helv.\ Phys.\ Acta \textbf{63} (1990) 3.
%
\bibitem{fri04}M.\ Frink, U.\ G.\ Meissner, J.\ High Energy Phys.\
  \textbf{0407} (2004) 028. 
%
\bibitem{oll06a}J.\ A.\ Oller, J.\ Prades, M.\ Verbeni, J.\ High Energy
  Phys.\  \textbf{0609} (2006) 079,\\ Erratum: {arxiv:hep-ph/0701096}.
%
\bibitem{clo93}F.\ E.\ Close, R.\ G.\ Roberts, Phys.\ Lett.\ B
  \textbf{316} (1993) 165.
%
\bibitem{bor99}B.\ Borasoy, Phys.\ Rev.\ D \textbf{59} (1999) 054021. 
%
\bibitem{rat99}P.\ G.\ Ratcliffe, Phys.\ Rev.\ D \textbf{59} (1999)
  014038.
%
\bibitem{oll01}J.\ A.\ Oller, U.\ G.\ Meissner, Phys.\ Lett.\ B
  \textbf{500} (2001) 263.
%
\bibitem{ose98}E.\ Oset, A.\ Ramos, Nucl.\ Phys.\ A \textbf{635}
  (1998) 99.
%
\bibitem{jid02}D.\ Jido, E.\ Oset, A.\ Ramos, Phys.\ Rev.\ C
  \textbf{66} (2002) 055203.
%
\bibitem{col65} S.\ Coleman, Nuov.\ Cim.\ \textbf{38} (1965) 438.
%
\bibitem{ben02}E.\ Oset, A.\ Ramos, C.\ Bennhold, Phys.\ Lett.\ B
  \textbf{527} (2002) 99, Erratum:  Phys.\ Lett.\ B \textbf{530}
  (2002) 260. 
%
\bibitem{mei00} U.\ G.\ Meissner, J.\ A.\ Oller, Nucl.\ Phys.\ A
  \textbf{673} (2000) 311.
%
\bibitem{lee94} C.-H.\ Lee, H.\ Jung, D.-P.\ Min, M.\ Rho, Phys.\
  Lett.\ B \textbf{326} (1994) 14.
%
\bibitem{tho79}G.\ 't Hooft, M.\ Veltman, Nucl.\ Phys.\ B
  \textbf{153} (1979) 365.
%
\bibitem{den91}A.\ Denner, U.\ Nierste, R.\ Scharf, Nucl.\ Phys.\ B
  \textbf{367} (1991) 637.
%
\bibitem{pas79}G.\ Passarino, M.\ Veltman, Nucl.\ Phys.\ B
  \textbf{160} (1979) 151.
%
\bibitem{sch70}H.\ Schweinler, E.\ Wigner, J.\ Math.\ Phys.\
  \textbf{11} (1970) 1693.
\end{thebibliography}
\end{document}